\RequirePackage{fix-cm}
\documentclass[twocolumn,epjc3]{svjour3}
\smartqed
\DeclareUnicodeCharacter{2032}{\ensuremath{^{\prime}}}
\usepackage{graphics}
\usepackage[dvipsnames]{xcolor}
\usepackage{graphicx,color}
\usepackage{enumitem}
\usepackage{cite}
\usepackage{braket}
\usepackage{slashed}
\usepackage[utf8]{inputenc}
\usepackage[
    colorlinks=true,
    linkcolor=blue,
    citecolor=blue,
    urlcolor=blue
]{hyperref}
\usepackage{textcomp}
\usepackage{float}
\usepackage{makecell}
\usepackage{epsfig}
\usepackage{amsmath,amssymb}
\usepackage[normalem]{ulem}
\usepackage{tikz}
\newcommand{\nn}{\nonumber}
\renewcommand{\t}{\tilde}
\newcommand{\h}{\hat}
\usetikzlibrary{shapes.geometric, arrows}
\usepackage{placeins}
\usepackage{microtype}
\hypersetup{hypertexnames=false}

\makeatletter
\renewcommand{\@makefntext}[1]{\noindent\@makefnmark #1}
\makeatother

\usepackage{etoolbox}
\makeatletter
\renewcommand{\makeheadbox}{}
\patchcmd{\@maketitle}{\hrule\@height0.35mm\noindent}{\noindent}{}
  {\PackageError{arxiv-layout}{Could not remove the journal title rule}{}}
\makeatother
\journalname{}

\begin{document}
\title{Big Crunching Your Big Bang: Doubly Warped Time-Dependent Type II Flux Backgrounds and Scale Separation}

\author{Kazi Fahim Reza Arko\thanksref{e1,addr1}
\and Walid Hasan\thanksref{e2,addr1,addr2}
\and M. Fahim Hoque\thanksref{e3,addr1}
\and Ahmed Rakin Kamal\thanksref{e4,addr3}
}

\institute{Department of Mathematics and Physical Sciences,
BRAC University, Kha 224, Bir Uttam Rafiqul Islam Avenue,
Dhaka 1212, Bangladesh \label{addr1}
\and
Department of Physics \& Astronomy, University of Manitoba, Winnipeg, Manitoba R3T 2N2, Canada \label{addr2}
\and
Department of Theoretical Physics and Astrophysics, Faculty of Science, Masaryk University, Kotlářská 2, CS-61137 Brno, Czechia \label{addr3}
}

\thankstext{e1}{e-mail: fahim.reza.arko@g.bracu.ac.bd}
\thankstext{e2}{e-mail: hasanw@myumanitoba.ca}
\thankstext{e3}{e-mail: fahim.hoque@g.bracu.ac.bd}
\thankstext{e4}{e-mail: ahmedrakinkamaltunok@gmail.com}

\date{}

\maketitle

\begin{abstract}
We construct three analytic, doubly warped cosmological solutions of massless type IIA supergravity with different choices of fluxes, namely, $H_3$, $\t F_2$ and $\t F_4$ along internal directions. The $H_3$ solution also belongs to the common NS--NS sector of type IIB.  Our solutions show that, without localised sources, nonconstant warping on a smooth compact space is possible. We derive their eleven-dimensional uplifts and four-dimensional Einstein-frame cosmology. Each universe undergoes (decelerating) expansion from a Big Bang, reaches a maximum size, and undergoes contraction towards a Big Crunch. We also analyse the scale separation of these backgrounds and find that they exhibit a small Einstein frame Hubble scale to KK mass ratio $\frac{|H_E|}{m_{\mathrm{KK}}}\ll1,$ although for a brief period of time, around the maxima of the scale factor in the Einstein frame. Finally, we analyse the supersymmetry of these backgrounds, which demonstrates that all three solutions break all forms of type IIA supersymmetry.
\end{abstract}

\section{Introduction}
\label{sec:introduction}

The large hierarchy between the electroweak and Planck scales is one of the basic motivations for considering extra dimensions. Randall and Sundrum showed that a warped extra dimension can generate this hierarchy through a warped compactification \cite{RandallSundrum1999}. In their construction, a mass parameter on the brane is related to the fundamental scale by $m_{\mathrm{IR}}=e^{-k\pi r_c}m_{\mathrm{UV}}$, where $k$ is the curvature scale and $r_c$ is the compactification radius. For a fixed radius, an exponential hierarchy therefore follows from $kr_c$. This raises the question of how such warped geometries may arise in supergravity and string theory.

In type IIB theory, the Klebanov--Strassler solution realises a smooth warped throat supported by fluxes on the deformed conifold \cite{KlebanovStrassler2000}. The Giddings--Kachru--Polchinski construction embeds such throats into compact flux backgrounds, combining a warped hierarchy of scales with partial moduli stabilisation \cite{Giddings:2001yu}. This framework provides a starting point for complete moduli stabilisation proposals (alongside de Sitter (dS) vacua), including KKLT \cite{Kachru:2003aw}, the Large Volume Scenario (LVS) \cite{Balasubramanian:2005zx}, and their perturbative cousins \cite{Cicoli:2024bwq, Chakraborty:2025wqn}.

Alongside the search for dS vacua, time-dependent compactifications provide a direct setting for studying dynamical 4D cosmology. The dS swampland conjectures have sharpened interest in this direction by questioning the existence of controlled, metastable dS vacua \cite{Obied:2018sgi,Ooguri:2018wrx}. There is also an observational motivation from the DESI DR2 measurements, combined with cosmic microwave background and supernova data, which show a preference for an evolving dark-energy equation of state \cite{DESI2025DR2}. An important point to note is that these results do not establish a dynamical dark energy, but do strengthen the motivation for investigating cosmologies beyond the classic dS description.

Time dependence also changes the assumptions entering classical compactification no-go theorems, including the Maldacena--Nu\~nez argument \cite{MaldacenaNunez2001}. Townsend and Wohlfarth demonstrated transient accelerated expansion from a compact negatively curved internal space with an evolving volume \cite{TownsendWohlfarth2003}. Related time-dependent supergravity backgrounds were constructed in the S-brane framework \cite{Chen:2002yq} alongside their compactifications \cite{Ohta:2003pu, Emparan:2003gg}. Examples of time-dependent compactifications and possible constraints have been studied in Refs.~\cite{Mueller:1989in, Bhattacharya:2003sh, Faruk:2026qwb, Buchel:2002kj, Sabra:2020gio, Tonioni:2026gmq}. More general accelerating cosmologies in type IIA supergravity were constructed by Marconnet and Tsimpis \cite{MarconnetTsimpis2023}, while the role of negative external spatial curvature was studied in \cite{AndriotTsimpisWrase2023}. More recently, ref. \cite{Kamal:2025qia} constructed time-dependent type IIB flux backgrounds without localised sources. They also examined how an evolving internal volume and flux components carrying one, two, or three external indices modify the traced Einstein equations, so that the usual no-go argument alone no longer excludes a positive four-dimensional Einstein-frame Ricci scalar.

Combining internal space warping with time dependence requires solving more challenging supergravity equations (as will be shown later). In particular, the warp factor, internal volume, and dilaton cannot generally be varied independently. Studies of warped moduli dynamics and the universal breathing mode make these constraints explicit \cite{GiddingsMaharana2006,FreyEtAl2009,Underwood2011}. This motivates the construction of explicit higher-dimensional solutions in which the internal geometry and cosmological evolution are determined together in hopes of solving the hierarchy problem.

A related question is whether the resulting solution admits a four-dimensional description separated from the massive Kaluza--Klein (KK) modes. For a vacuum with curvature radius $L_4$, scale separation requires $m_{\mathrm{KK}}L_4\gg1$. Parametric scale separation means that this hierarchy can be made arbitrarily large in a controlled family of solutions. The DGKT construction serves as a prominent candidate in massive type IIA theory, where a large-flux limit gives weak coupling, large internal volume, and a hierarchy between the KK and AdS scales \cite{DeWolfeEtAl2005}. Non-supersymmetric scale-separated AdS solutions have also been found \cite{MarchesanoQuirant2019}, although their stability is a separate question \cite{MarchesanoQuirantZatti2022}.

These constructions are closely connected to the AdS distance conjecture \cite{LustPaltiVafa2019}. Its general form requires a tower of states to become light as the cosmological constant approaches zero. The strong supersymmetric version further requires the tower scale to track the AdS curvature scale, thereby excluding parametric separation from that tower. It is this stronger statement that is challenged by supersymmetric scale-separated candidates. Holographic consistency provides another way to investigate these candidates \cite{Bobev:2025yxp, Apers:2026lgi}. For time dependent backgrounds, Andriot, Cribiori and Van Riet showed that classical flux-supported cosmologies can develop a hierarchy between the Hubble and KK scales without orientifolds \cite{Andriot:2025cyi}. In a spatially flat cosmology, $H_E$ needs to satisfy the hierarchy
\begin{equation}
    \frac{|H_E|}{m_{\mathrm{KK}}}\ll1,
    \label{eq:intro-curvature-hierarchy}
\end{equation}
with all quantities evaluated in the same four-dimensional Einstein frame.

In this paper, we construct three analytic cosmological backgrounds with smooth compact internal spaces and no localised sources, unlike the GKP construction that needs localised sources. We use the doubly warped metric in \eqref{eq:CommonAnsatz}, in which the external metric carries an internal warp factor $A(y)$ and the internal metric carries a time-dependent scale factor $\beta(t)$. For a single internal flux and an additively separated dilaton as a function of internal space and time, the mixed Einstein equation relates the dilaton to these two functions, while the integrability condition on the compact internal manifold fixes their relative scaling. The three solutions are supported by $H_3$, $\t F_2$ and $\t F_4$, respectively. Their internal spaces are a deformed $S^3\times T^3$, an $S^2$ fibred over $S^2$ times a further $S^2$, and an $S^4$ fibred over a compact hyperbolic surface. All three are solutions of massless type IIA supergravity, and the $H_3$ background also belongs to the common NS--NS sector of type IIB. We give their eleven-dimensional uplifts as well. 

The four-dimensional Einstein-frame cosmologies expand from a big bang, reach a maximum size, and contract towards a big crunch. Their expansion is decelerating. We determine the separated internal eigenfunctions and mass gaps for massless minimally coupled scalars. These calculations allow us to compare the cosmological scales with the KK spectrum. Although $|H_E|/m_{\mathrm{KK}}$ becomes arbitrarily small near turnaround, $|dH_E/dT_E|/m_{\mathrm{KK}}^2$ remains close to order one. Thus, these backgrounds exhibit a small Hubble-to-KK hierarchy for a very brief period of time. We also observe that all three backgrounds break type IIA supersymmetry.

The paper is organised as follows. Section~\ref{sec:ansatz-eoms} presents the ansatz, construction strategy, and M-theory uplifts. Section~\ref{sec:compact-backgrounds} gives the explicit solutions and their cosmological evolution. Section~\ref{sec:scale-separation} studies the KK masses and scale hierarchies. Section~\ref{sec:supersymmetry} discusses supersymmetry, and Section~\ref{sec:outlook} discusses possible extensions towards accelerated expansion. Our conventions, detailed field equations and internal manifold eigenvalue calculations are collected in~\ref{app:typeii-conventions}, \ref{app:detailed-eoms} and~\ref{app:kk-spectrum}, respectively.
\section{Ansatz, general strategy and M-theory uplift of the compact solutions}
\label{sec:ansatz-eoms}

We work in ten-dimensional Einstein frame and follow the conventions of \cite{Kamal:2025qia}, collected in~\ref{app:typeii-conventions}. Our metric and flux ansatz is
\begin{align}
 ds_{10}^2&=e^{2A(y)}\h g_{\mu\nu}(t)dx^\mu dx^\nu+e^{2\beta(t)}\t g_{mn}(y)dy^m dy^n,\nn\\
 \text{IIB}:\quad
 \t F_5&=\Omega(x)\t d\xi(y)\wedge\h\epsilon-\Omega(x)e^{4\beta-4A}\t\star\t d\xi(y),\nn\\
 G_3&=\alpha(x)\delta_3(y),\nn\\
 \text{IIA}:\quad
 F_0&=m,\quad H_3=h_3(y),\quad\t F_2=\delta_2(y),\nn\\
 \t F_4&=\delta_4(y)+\varpi(x)\h\epsilon.
 \label{eq:CommonAnsatz}
\end{align}
The IIA and IIB fluxes have the same metric ansatz. Hats and tildes refer to the unwarped metrics $\h g$ and $\t g$. The flux components are real except for $\alpha(x)$, which may be complex, and we retain $\tau(x,y)$ with $\operatorname{Im}\tau>0$ in IIB and a real $\Phi(x,y)$ in IIA. The five-form is self-dual by construction. The Bianchi identities, form equations, and Einstein sources are given in~\ref{app:detailed-eoms}.

\subsection{General strategy for the solutions}
We take the internal space to be smooth, compact and without boundary, and have no localised sources. The compact solutions in Section~\ref{sec:compact-backgrounds} carry a single flux with legs in the internal space, $H_3$, $\t F_2$ or $\t F_4$, with all other fluxes zero. The $H_3$ solution belongs to the common NS--NS sector of IIA and IIB. In IIB notation, $C_0=F_3=\t F_5=0$ and $G_3=-ie^{-\Phi}H_3$. For these choices of fluxes, the mixed flux energy-momentum tensor vanishes. Taking both $A$ and $\beta$ nonconstant and assuming an additively separated dilaton, \eqref{eq:Ricci_mixed} and \eqref{eq:iia-Einstein-mixed} give
\begin{equation}
 \Phi=\kappa\beta+\frac{16}{\kappa}A,
 \qquad\kappa\neq0,
 \label{eq:strategy-dilaton-separation}
\end{equation}
Using $P_p$ and $a_p$ which are the form and the dilaton coupling of the p-forms, respectively, as defined in \eqref{eq:iia-internal-flux-notation}, the internal p-form flux carries the relative time dependence $e^{E_p\beta}$, with $E_p=a_p\kappa-2(p-1)$. Within these single-flux cases, nonconstant $A$ on a compact space requires $E_p=0$. \footnote{Combining the dilaton equation
with its integral over $M_6$ gives

\begin{align}
\label{eq:strategy-integrability}
 \t\nabla^{\t2}e^{4A}
 &=\frac{a_p\kappa}{8}e^{E_p\beta}
 \Biggl[e^{(4+16a_p/\kappa)A}(P_p\cdot P_p) \\\nn 
 &-e^{2A}
 \frac{\int_{M_6}e^{(4+16a_p/\kappa)A}(P_p\cdot P_p)\,\t\epsilon}
      {\int_{M_6}e^{2A}\,\t\epsilon}\Biggr],
\end{align}

where the form contraction uses $\t g_{mn}$, and $\t\epsilon$ is its volume form. Both the left-hand side and the bracket depend only on $y$. If $E_p\neq0$ and $\beta$ is nonconstant, the time-dependent factor forces the bracket to vanish.} Without this, compactness would force $e^{4A}$ to be constant. This selects
\begin{equation}
 \begin{array}{c|ccc}
 \text{flux} & H_3 & \t F_2 & \t F_4\\ \hline
 a_p & -1 & 3/2 & 1/2\\
 \kappa & -4 & 4/3 & 12
 \end{array}
 \label{eq:strategy-compact-branches}
\end{equation}
The external scalar combination and the internal warp equation then separate with a constant $X_0$:
\begin{align}
 \h\nabla^{\h2}\beta
 +6\partial_\mu\beta\partial^{\h\mu}\beta
 &=X_0e^{-2\beta},\nn\\
 \t\nabla^{\t2}e^{4A}
 &=\frac14\Bigl[
 (p-1)e^{(4+16a_p/\kappa)A}(P_p\cdot P_p)\nn\\
 &\hspace{18mm}-\kappa^2X_0e^{2A}\Bigr].
 \label{eq:strategy-separated-warp}
\end{align}
We solve the internal form and Einstein equations subject to this condition, and then solve the remaining time equations with $\alpha_1=\alpha_2=\alpha_3$. Smoothness, flux quantisation and the four-dimensional Einstein-frame cosmology are discussed with each solution in Section~\ref{sec:compact-backgrounds}.

\subsection{M-theory uplift}
All three solutions have $m=0$ and admit the massless IIA uplift \cite{Becker:2006dvp, Polchinski:1998rr}
\begin{align}
 ds_{11}^2&=e^{-\Phi/6}ds_{10}^2
             +e^{4\Phi/3}(dz+C_1)^2,\nn\\
 G_4&=\t F_4+H_3\wedge(dz+C_1),\qquad dC_1=\t F_2.
 \label{eq:strategy-Mtheory-uplift}
\end{align}
The combination $dz+C_1$ is the M-theory circle. For the $H_3$ solution, we first regard the common NS--NS fields as a IIA background. Taking $C_1=0$ and $\Phi=-4A-4\beta$ gives
\begin{align}
 ds_{11}^2&=e^{(8A+2\beta)/3}\h ds^2
          +e^{(2A+8\beta)/3}\t ds^2+e^{-16(A+\beta)/3}dz^2,\nn\\
 G_4&=h_3\wedge dz.
 \label{eq:strategy-H3-uplift}
\end{align}
For periodic $z$, the internal topology is $S^3\times T^4$. The circle radius depends on both $t$ and $y$, so the uplift remains warped.

For the two-form solution, $\Phi=4\beta/3+12A$ gives
\begin{align}
 ds_{11}^2&=e^{-2\beta/9}\h ds^2+e^{16\beta/9}\t ds_7^2,
 \qquad G_4=0,\nn\\
 ds_7^2&=e^{-2A} ds^2+e^{16A}(dz+C_1)^2,\nn\\
 \t R(g_7)&=\frac{56 X_0}{9} = \frac{7}{L^2}.
 \label{eq:strategy-F2-uplift}
\end{align}
Thus the uplift is an eleven-dimensional vacuum cosmology with a time-dependent internal scale factor and no internal-coordinate dependence in the external metric factor. For our explicit solution, $g_7$ is the product metric on $S^3(\sqrt2L)\times S^2(L)\times S^2(L)$. The ten-dimensional warp factor is generated by reducing along a circle whose norm varies over this space.

For the four-form solution, $C_1=0$ and $\Phi=12\beta+4A/3$ give
\begin{align}
 ds_{11}^2&=e^{16A/9}ds_5^2+e^{-2A/9}\t ds^2,
 \qquad G_4=\delta_4,\nn\\
 ds_5^2&=e^{-2\beta}\h ds^2+e^{16\beta}dz^2,\nn\\
 \t R(g_5)&=-40 X_0 = - \frac{10}{L^2}.
 \label{eq:strategy-F4-uplift}
\end{align}
Our example uses the $S^4$ fibred over $\Sigma_{g\geq2}$ in the solution of Maldacena and Nu\~nez \cite{MaldacenaNunez2001}.
\section{Explicit construction of the backgrounds}
\label{sec:compact-backgrounds}

We now construct the three compact backgrounds described in Section~\ref{sec:ansatz-eoms}, supported respectively by $H_3$, $\t F_2$ and $\t F_4$. The first belongs to the common NS--NS sector of IIA and IIB; the other two are IIA solutions. We impose spatial isotropy and retain the conventions of~\ref{app:typeii-conventions}. All three internal manifolds are smooth, compact and without boundary, and no localised sources are introduced.

For the cosmological interpretation, the coefficient of $\h R$ in the dimensionally reduced action is
\begin{align}
 S_{\rm grav}\supset
 \frac{\mathcal V_A}{2\kappa_{10}^2}
 \int d^4x\sqrt{-\h g}\,e^{6\beta}\h R\,,
 \qquad
 \mathcal V_A=\int_{M_6}e^{2A}\t\epsilon\,.
 \label{eq:CompactEinsteinReduction}
\end{align}
Thus the internal warp factor contributes to internal manifold volume $\mathcal V_A$. Since $A$ and $\t g$ are time-independent, this integral is constant and can be absorbed into $M_4^2=\mathcal V_A/\kappa_{10}^2$. The four-dimensional Einstein metric is therefore
\begin{align}
 ds_{4E}^2&=e^{6\beta}\h ds^2
       =-dT_E^2+a_E^2\sum_{i=1}^3dx^idx^i\,,\nn\\
 dT_E&=e^{\sigma/2+3\beta}dt\,,
 \qquad a_E=e^{3\beta+\alpha_1/2}\,.
 \label{eq:CompactEinsteinMetric}
\end{align}
In the last line of the above equation, expansion and acceleration refer respectively to the signs of $da_E/dT_E$ and $d^2a_E/dT_E^2$.

\subsection{Background with non-trivial $G_3$}
\label{subsec:compact-G3}

We first consider the case
\begin{align}
 \t F_5&=0\,,\qquad G_3=\alpha(t)\delta_3(y)\,,
 \qquad \alpha=-ie^{4\beta}\,,\nn\\
 \operatorname{Re}\tau &=0\,,\qquad
 \operatorname{Im}\tau=e^{4A+4\beta}\,,
 \label{eq:ARKTrial3Branch}
\end{align}
where $\delta_3$ is real. Since $G_3\wedge\bar G_3=0$, the five-form Bianchi identity in \eqref{eq:iib-complex-system} is satisfied. Equations~\eqref{eq:G3Bianchi13}, \eqref{eq:G3Bianchi04} and \eqref{eq:G3EOM44} reduce to
\begin{align}
 \t d\delta_3=4\t dA\wedge\delta_3\,,
 \qquad
 \t d\bigl(e^{4A}\t\star\delta_3\bigr)=0\,.
 \label{eq:ARKTrial3FluxConstraints}
\end{align}
The physical fields of type IIB are then
\begin{align}
 C_0=F_3=0\,,\quad
 \Phi=-4A-4\beta\,,\quad H_3= h_3 =e^{-4A}\delta_3\,.
 \label{eq:ARKTrial3PhysicalFields}
\end{align}
The same metric, dilaton, and $H_3$ solve the IIA equations with $m=\t F_2=\t F_4=0$. The mixed Einstein equation is satisfied because
\begin{align}
 \frac{\partial_{(\mu}\tau\partial_{m)}\bar\tau}
      {2(\operatorname{Im}\tau)^2}
 =8\partial_\mu\beta\partial_mA=R_{\mu m}\,.
 \label{eq:ARKTrial3MixedCheck}
\end{align}
The axiodilaton equation \eqref{eq:tau-components-reduced} separates as
\begin{align}
 &e^{2\beta}\bigl(\h\nabla^{\h2}\beta
       +6\partial_\mu\beta\partial^{\h\mu}\beta\bigr)=C\,,
       \nn\\
 &\t\nabla^{\t2}e^{4A}
       =\frac12\delta_3\cdot\delta_3-4Ce^{2A}\,.
 \label{eq:ARKTrial3SeparatedTau}
\end{align}
Here, $C=X_0>0$ is consistent with \eqref{eq:strategy-separated-warp}.

\subsubsection{Internal manifold}
We use coordinates $y^m=(\theta,\varphi_1,\varphi_2,z_1,z_2,z_3)$, with $0\leq\theta\leq\pi/2$ and period $2\pi$ for $\varphi_1,\varphi_2,z_a$. We take the metric to be
\begin{align}
 \t ds^2=e^{2A(\theta)}\Bigl[
 L^2\bigl(d\theta^2+f^2d\varphi_1^2+g^2d\varphi_2^2\bigr)
       +\sum_{a=1}^3R_a^2dz_a^2\Bigr]\,,
 \label{eq:ARKTrial3TildedMetric}
\end{align}
where $L,R_a>0$. With orientation $d\theta\wedge d\varphi_1\wedge d\varphi_2 \wedge dz_1\wedge dz_2\wedge dz_3$, the form equations are solved by
\begin{align}
 \delta_3&=QL^3fg e^{-4A}
             d\theta\wedge d\varphi_1\wedge d\varphi_2\,,
             \nn\\
 e^{4A}\t\star\delta_3
       &=-QR_1R_2R_3dz_1\wedge dz_2\wedge dz_3\,, \nn \\
 \delta_3\cdot\delta_3 &=Q^2e^{-14A}\,.
 \label{eq:ARKTrial3FluxSolution}
\end{align}
The minus sign follows from \eqref{eq:hodge-convention}. The internal Einstein and internal $\tau$ equations become
\begin{align}
 &-\frac{f''}{f}-\frac{g''}{g}-8A''
       =\frac{L^2Q^2}{2}e^{-16A}\,,\nn\\
 &-\frac{f''}{f}-\frac{f'g'}{fg}
       -8\frac{f'}fA'
       =\frac{L^2Q^2}{2}e^{-16A}\,,\nn\\
 &-\frac{g''}{g}-\frac{f'g'}{fg}
       -8\frac{g'}gA'
       =\frac{L^2Q^2}{2}e^{-16A}\,,\nn\\
 &\frac1{L^2}\left[
A''+\left(\frac{f'}f+\frac{g'}g\right)A'+8(A')^2
 \right]-\frac{Q^2}{8}e^{-16A}=-C\,.
 \label{eq:ARKTrial3InternalODEs}
\end{align}
Here, primes denote $\theta$-derivatives. The torus equation coincides with the last equation, i.e the internal $\tau$ equation and the off-diagonal components vanish. These equations admit the solution
\begin{align}
 A&=\frac18\ln\bigl(\cos^2\theta+c_1^2\sin^2\theta\bigr)\,,
       \nn\\
 f&=\frac{\sin\theta}
          {\sqrt{\cos^2\theta+c_1^2\sin^2\theta}}\,,
 \qquad
 g=\frac{c_1\cos\theta}
          {\sqrt{\cos^2\theta+c_1^2\sin^2\theta}}\,,
       \nn\\
 Q&=\frac{2c_1}{L}\,,\qquad
 C=\frac1{2L^2}\,,\qquad c_1>0\,.
 \label{eq:ARKTrial3InternalSolution}
\end{align}
The $H_3$ flux is therefore
\begin{align}
 H_3=
 \frac{2L^2c_1^2\sin\theta\cos\theta}
      {(\cos^2\theta+c_1^2\sin^2\theta)^2}
       d\theta\wedge d\varphi_1\wedge d\varphi_2\,.
 \label{eq:ARKTrial3FinalPhysicalFlux}
\end{align}

For every finite $c_1>0$, the internal geometry is smooth. The $\varphi_1$ circle closes at $\theta=0$, where $f\sim\theta$, and the $\varphi_2$ circle closes at $\theta=\pi/2$, where $g\sim\pi/2-\theta$. Neither endpoint is a boundary. The topology is $S^3\times T^3$, and $c_1=1$ gives the round sphere with $A=0$. We take $c_1\neq1$ for a nonconstant internal warp factor $A$. The integrated warp equation holds as
\begin{align}
 \frac12\int_{S^3\times T^3}
       \t\epsilon\,\delta_3\cdot\delta_3
 &=4C\int_{S^3\times T^3}\t\epsilon\,e^{2A}\nn\\
 &=(2\pi)^5c_1LR_1R_2R_3\,.
 \label{eq:ARKTrial3Integrability}
\end{align}
Both sides contain the same factor of $c_1$. Moreover, unlike the time-dependent backgrounds in \cite{Kamal:2025qia}, the flux quantisation also goes through
\begin{align}
 \int_{S^3}H_3=4\pi^2L^2\,,
 \qquad \frac{L^2}{\alpha'}\in\mathbb Z_{>0}\,.
 \label{eq:ARKTrial3FluxQuantization}
\end{align}
Also, flux quantisation does not fix $c_1$.

\subsubsection{Isotropic external space solution}
We take
\begin{align}
 \h ds^2&=-e^\sigma dt^2
              +e^{\alpha_1}\sum_{i=1}^3dx^idx^i,
 \label{eq:ARKTrial3HattedMetric}
\end{align}
with $\sigma=2\beta$, and $\alpha_1=\alpha_2=\alpha_3$.
Here $\alpha_i$ are metric functions, distinct from the three-form coefficient $\alpha$. The remaining equations are
\begin{align}
 \ddot\beta+
 \left(\frac32\dot\alpha_1+5\dot\beta\right)\dot\beta
       &=-\frac1{2L^2}\,,\nn\\
 \frac12\ddot\alpha_1-\ddot\beta
 +\left(\frac12\dot\alpha_1-\dot\beta\right)
  \left(\frac32\dot\alpha_1+5\dot\beta\right)&=0\,,\nn\\
 3\left[
 \frac12\ddot\alpha_1-\ddot\beta
 +\left(\frac12\dot\alpha_1-\dot\beta\right)^2
 \right]+8\ddot\beta &=0\,.
 \label{eq:ARKTrial3ExternalODEs}
\end{align}
Dots denote $t$-derivatives. Choosing the case with increasing spatial scale of $\h g_{\mu\nu}$ gives
\begin{align}
 \beta&=-\frac{\Phi_0}{4}
       +\frac14\ln\cos(t/L)
       +\frac{1-\sqrt3}{8}\ln\tan(t/L)\,,\nn\\
 \alpha_1&=\alpha_2=\alpha_3
       =2\beta+\frac2{\sqrt3}\ln\tan(t/L)\,,
 \quad 0<t<\frac{\pi L}{2}\,.
 \label{eq:ARKTrial3TimeSolution}
\end{align}
Here $\Phi_0$ is real. Throughout this interval, $\dot\beta<0$ and $\dot\alpha_i>0$: the internal scale decreases while the hatted spatial scale increases.

\subsubsection{Four-dimensional Einstein-frame cosmology}
Einstein-frame equation~\eqref{eq:CompactEinsteinMetric} gives us
\begin{align}
 dT_E&=e^{4\beta}dt\,,\nn\\
 a_E&=e^{-\Phi_0}\cos(t/L)
       \bigl[\tan(t/L)\bigr]^{(1-1/\sqrt3)/2}\,.
 \label{eq:ARKTrial3EinsteinMetric}
\end{align}
Its proper-time derivatives are
\begin{align}
 \frac{da_E}{dT_E}
 &=\frac{a_Ee^{-4\beta}}{L\sin(2t/L)}
       \left[\cos(2t/L)-\frac1{\sqrt3}\right]\,,
 \label{eq:ARKTrial3EinsteinExpansion}\\
 \frac{d^2a_E}{dT_E^2}
 &=\frac{4a_Ee^{-8\beta}}{3L^2\sin^2(2t/L)}
       \left[\sqrt3\cos(2t/L)-2\right]<0\,.
 \label{eq:ARKTrial3EinsteinAcceleration}
\end{align}
The Ricci scalar is
\begin{align}
 R_{4E}
 &=\frac{2e^{-8\beta}}{L^2\sin^2(2t/L)}
 \Bigl[3\cos^2(2t/L)\nn\\
 &\hspace{25mm}
       +2\sqrt3\cos(2t/L)-7\Bigr]<0\,.
 \label{eq:ARKTrial3EinsteinRicci}
\end{align}
The Einstein-frame universe expands for $0<t<t_*$, where $t_*=\frac L2\arccos(1/\sqrt3)$. It reaches its maximum size at $t=t_*$ and subsequently contracts. The expanding phase is decelerating, and during contraction $|da_E/dT_E|$ increases. There is no accelerated expansion. We show the plots for this in Figure~\ref{fig:einstein-frame-cosmology}. 

\subsubsection{String-frame interpretation}
Using $g^{(s)}_{MN}=e^{\Phi/2}g_{MN}$, we find
\begin{align}
 ds_{10,s}^2={}&-dt^2+
       \bigl[\tan(t/L)\bigr]^{2/\sqrt3}
                         \sum_{i=1}^3dx^idx^i\nn\\
 &+L^2\bigl(d\theta^2+f^2d\varphi_1^2
                             +g^2d\varphi_2^2\bigr)
       +\sum_{a=1}^3R_a^2dz_a^2\,.
 \label{eq:ARKTrial3StringFrame}
\end{align}
Interestingly, the string-frame internal metric is static, while $\Phi=-4A-4\beta$ depends on both $t$ and $\theta$. Furthermore, its internal metric and flux belong to the marginally deformed $SU(2)$ WZW family \cite{HassanSen1993,Hoare2022,AliajSfetsosSiampos2023}. This identification concerns the internal background, and thus the complete time-dependent solution has been constructed at the level of the supergravity equations, whose CFT story remains unaddressed, unlike the time-independent constructions. Its M-theory uplift is given in \eqref{eq:strategy-H3-uplift}.

\subsection{Background with non-trivial $\t F_2$}
\label{subsec:compact-F2}

We next consider the $\kappa=4/3$ IIA case of \eqref{eq:strategy-compact-branches},
\begin{align}
 m=H_3=\t F_4=0\,,\quad
 \t F_2=\delta_2(y)\,,\quad
 \Phi=\frac43\beta+12A\,.
 \label{eq:CompactF2Branch}
\end{align}

\subsubsection{Internal manifold}
The internal space consists of an $S^2$ fibre $(\theta,\chi)$ over a base $S^2$ $(\theta_1,\varphi_1)$, times a further $S^2$ $(\theta_2,\varphi_2)$. The polar angles lie in $[0,\pi]$ and the azimuthal angles have period $2\pi$. Defining
\begin{align}
 D&=\cos^2\theta+c^2\sin^2\theta\,,
 \qquad c^2=1+\frac{p^2}{2}\,,\nn\\
 e^\chi&=d\chi-\frac p2\cos\theta_1d\varphi_1\,,
 \qquad A=\frac1{16}\ln D\,.
 \label{eq:CompactF2Definitions}
\end{align}
The one-form $e^\chi$ couples the fibre azimuthal angle to the base with integer $p$. The metric is
\begin{align}
 \t ds^2={}&e^{2A}L^2\Bigl[
 \frac12\bigl(d\theta_1^2+\sin^2\theta_1d\varphi_1^2\bigr)
       +d\theta^2+\frac{\sin^2\theta}{D}(e^\chi)^2\nn\\
 &\hspace{20mm}
       +d\theta_2^2+\sin^2\theta_2d\varphi_2^2\Bigr]\,,
 \qquad L>0\,.
 \label{eq:CompactF2Metric}
\end{align}
The $\chi$ circle closes smoothly at $\theta=0,\pi$, completing the $S^2$ fibre. On the northern and southern patches of the base, we have
\begin{align}
 \chi_N=\chi-\frac p2\varphi_1\,,
 \qquad \chi_S=\chi+\frac p2\varphi_1\,.
 \label{eq:CompactF2Patches}
\end{align}
The transition $\chi_S=\chi_N+p\varphi_1$ requires $p\in\mathbb Z$. We take $p\in2\mathbb Z\setminus\{0\}$, so the internal manifold is diffeomorphic to $S^2\times S^2\times S^2$, with nonconstant $A$. The mixed Einstein equation \eqref{eq:iia-Einstein-mixed} holds identically for \eqref{eq:CompactF2Branch}. Writing $v_i=\sin\theta_i\,d\theta_i\wedge d\varphi_i$, the flux and local gauge potential are
\begin{align}
 \delta_2&=-\frac QD v_1
       +\frac{Qp\sin2\theta}{D^2}d\theta\wedge e^\chi\,,
 \qquad Q=\frac L{\sqrt2}\,,\nn\\
 C_1&=\frac QD\bigl(
       \cos\theta_1d\varphi_1+p\sin^2\theta\,d\chi\bigr)\,,
 \qquad \delta_2=\t dC_1\,.
 \label{eq:CompactF2Flux}
\end{align}
Since $\delta_2=\t dC_1$ locally, the Bianchi identity holds. With orientation $v_1\wedge d\theta\wedge e^\chi\wedge v_2$,
\begin{align}
 e^{22A}\t\star\delta_2
 =QL^2\Bigl[-2\sin\theta\,d\theta\wedge e^\chi
                    +p\cos\theta\,v_1\Bigr]\wedge v_2\,.
 \label{eq:CompactF2Hodge}
\end{align}
Using $\t de^\chi=(p/2)v_1$ and $\t dv_i=0$, the two derivative terms cancel. Hence $\t d(e^{22A}\t\star\delta_2)=0$ satisfies the form equation in \eqref{eq:iia-single-flux-branches}. The internal terms in the dilaton equation \eqref{eq:iia-dilaton-reduced} are
\begin{align}
 e^{20A}\delta_2\cdot\delta_2
 &=\frac{2[c^2+(c^2-1)\cos^2\theta]}{L^2D^2}\,,\nn\\
 \t\nabla^{\t2}e^{4A}
 &=\frac{e^{2A}}{2L^2}
 \left[\frac{c^2+(c^2-1)\cos^2\theta}{D^2}-1\right]\,.
 \label{eq:CompactF2WarpCheck}
\end{align}
Substitution into \eqref{eq:strategy-separated-warp} fixes
\begin{align}
 X_0=\frac9{8L^2}\,.
 \label{eq:CompactF2SeparationConstant}
\end{align}
The internal Einstein equation \eqref{eq:iia-Einstein-internal} becomes
\begin{align}
 &\t R_{mn}-4\t\nabla_m\t\nabla_nA
       -76\partial_mA\partial_nA
       -X_0e^{-2A}\t g_{mn}\nn\\
 &\qquad=\frac12e^{18A}
 \left[(\delta_2\cdot\delta_2)_{mn}
       -\frac18\t g_{mn}\delta_2\cdot\delta_2\right]\,.
 \label{eq:CompactF2InternalEinstein}
\end{align}
One can directly substitute and verify that all Einstein equation components, including the off-diagonal ones, are solved. Note that we have used the contraction defined in \eqref{eq:form-tensor-contraction}. The integrated condition follows using $u=\cos\theta$,
\begin{align}
 \mathcal V_A&=32\pi^3L^6\,,\nn\\
 \int_{M_6}\t\epsilon\,\t\nabla^{\t2}e^{4A}
 &=8\pi^3L^4
 \left[\frac{u}{c^2-(c^2-1)u^2}-u\right]_{-1}^{1}
 =0\,.
 \label{eq:CompactF2Integrability}
\end{align}
Thus \eqref{eq:strategy-integrability} holds without boundary or localised contributions. The regular connection potentials are $C_{1,N}=C_1-Qd\varphi_1$ and $C_{1,S}=C_1+Qd\varphi_1$. The sections at $\theta=0,\pi$ have
\begin{align}
 \int_{S_0}\delta_2=\int_{S_\pi}\delta_2=-4\pi Q\,.
 \label{eq:CompactF2Periods}
\end{align}
The flux through the $S^2$ fibre vanishes, since at fixed base coordinates
\begin{align}
 \int_{S^2_{\rm fibre}}\delta_2
 =2\pi Qp\left[\frac{\sin^2\theta}{D}\right]_0^\pi=0\,.
\end{align}
Note that there can be local flux on the fibre while its net integral vanishes. The last $S^2$ also carries no flux, as $\delta_2$ has no legs along it.

\subsubsection{Isotropic external space solution}
Our 4D metric ansatz is
\begin{align}
 \h ds^2&=-e^\sigma dt^2
             +e^{\alpha_1}\sum_{i=1}^3dx^idx^i\,,
 \qquad \alpha_1=\alpha_2=\alpha_3\,,\nn\\
 \sigma&=3\alpha_1+12\beta\,.
 \label{eq:CompactF2HattedMetric}
\end{align}
The choice of $\sigma$ simplifies the equations. The temporal equation in \eqref{eq:strategy-separated-warp} is $-e^{-\sigma+2\beta}\ddot\beta=X_0$. Together with the spatial and $tt$ equations in \eqref{eq:iia-Einstein-external}, it gives
\begin{align}
 \ddot\beta&=-\frac9{8L^2}e^{3\alpha_1+10\beta}\,,
 \qquad \ddot\alpha_1=\frac29\ddot\beta\,,\nn\\
 0&=\frac32\dot\alpha_1^2
       +18\dot\alpha_1\dot\beta+\frac{262}{9}\dot\beta^2
       +\frac7{L^2}e^{3\alpha_1+10\beta}\,.
 \label{eq:CompactF2TimeEquations}
\end{align}
A solution of all three equations is
\begin{align}
 \beta&=\frac34\Phi_0-\frac3{16}\ln\cosh(kt)
                         -\frac{\sqrt{21}}{16}kt\,,\nn\\
 \alpha_1&=\alpha_2=\alpha_3
       =\frac29\beta+\frac{2\sqrt{21}}9kt+\alpha_0\,,
       \nn\\
 e^{3\alpha_0}&=\frac{k^2L^2}{6}e^{-8\Phi_0}\,, \quad k>0\,,\quad -\infty<t<\infty\,.
 \label{eq:CompactF2TimeSolution}
\end{align}
The dilaton is
\begin{align}
 \Phi=\Phi_0+\frac34\ln D
       -\frac14\ln\cosh(kt)-\frac{\sqrt{21}}{12}kt\,.
 \label{eq:CompactF2Dilaton}
\end{align}
Both warp factors are nonconstant. The internal scale decreases throughout, since $\dot\beta=-k[3\tanh(kt)+\sqrt{21}]/16<0$.

\subsubsection{Four-dimensional Einstein-frame cosmology}
The choice of $\sigma$ in \eqref{eq:CompactF2HattedMetric} gives $dT_E=a_E^3dt$, and
\begin{align}
 a_E=e^{7\Phi_0/3+\alpha_0/2}
       [\cosh(kt)]^{-7/12}e^{-\sqrt{21}kt/12}\,.
 \label{eq:CompactF2EinsteinScale}
\end{align}
Consequently,
\begin{align}
 \frac{da_E}{dT_E}
 &=-\frac{k}{12a_E^2}
       \bigl[7\tanh(kt)+\sqrt{21}\bigr]\,,\nn\\
 \frac{d^2a_E}{dT_E^2}
 &=-\frac{k^2}{a_E^5}\left[
       \frac7{12}\operatorname{sech}^2(kt)\right.\nn\\
 &\hspace{15mm}\left.
       +\frac{[7\tanh(kt)+\sqrt{21}]^2}{72}\right]<0\,,
       \nn\\
 R_{4E}
 &=-\frac{k^2}{a_E^6}\left[
       \frac72\operatorname{sech}^2(kt)\right.\nn\\
 &\hspace{15mm}\left.
       +\frac{[7\tanh(kt)+\sqrt{21}]^2}{24}\right]<0\,.
 \label{eq:CompactF2Cosmology}
\end{align}
The universe expands for $\tanh(kt)<-\sqrt{3/7}$, reaches its maximum size at equality and then contracts. The expansion is decelerating. Although the hatted spatial scale factor increases throughout, the decreasing internal volume causes the Einstein-frame recollapse. We show the plots for this in Figure~\ref{fig:einstein-frame-cosmology}. 
\subsection{Background with non-trivial $\t F_4$}
\label{subsec:compact-F4}

Finally, we consider the $\kappa=12$ case of \eqref{eq:strategy-compact-branches},
\begin{align}
 m=H_3=\t F_2=0\,,\quad
 \t F_4=\delta_4(y)\,,\quad
 \Phi=12\beta+\frac43A\,.
 \label{eq:CompactF4Branch}
\end{align}

\subsubsection{Internal manifold}
The internal space is an $S^4$ fibred over a compact Riemann surface $\Sigma_g$ of genus $g\geq2$. The base has metric $ds_\Sigma^2$ of Gaussian curvature $-1$ and volume form $\epsilon_\Sigma$. Locally,
\begin{align}
 ds_\Sigma^2&=\frac{du_1^2+du_2^2}{u_2^2}\,,
 \qquad V=-\frac{du_1}{u_2}\,,
 \qquad \t dV=-\epsilon_\Sigma\,.
 \label{eq:CompactF4Base}
\end{align}
The fibre coordinates are $(\theta,\chi,\vartheta,\varphi)$, with $0\leq\theta\leq\pi/2$, $\chi$ of period $2\pi$, and $(\vartheta,\varphi)$ parametrising a unit $S^2$. The connection $V$ twists the $\chi$ circle over the base. Under $V\to V+\t d\Lambda$ and $\chi \to \chi-\Lambda$, the metric is invariant. Defining
\begin{align}
 W&=1+\cos^2\theta\,,
 \qquad A=\frac3{16}\ln W\,,\nn\\
 \epsilon_{S^2}&=\sin\vartheta\,d\vartheta\wedge d\varphi\,.
 \label{eq:CompactF4Definitions}
\end{align}
The internal metric is
\begin{align}
 \t ds^2={}&e^{2A}L^2\Bigl[
 ds_\Sigma^2+d\theta^2+\frac{\cos^2\theta}{W}
       \bigl(d\vartheta^2+\sin^2\vartheta\,d\varphi^2\bigr)
       \nn\\
 &\quad
 +\frac{2\sin^2\theta}{W}(d\chi+V)^2\Bigr]\,,
 \qquad L>0\,.
 \label{eq:CompactF4Metric}
\end{align}
For $0<\theta<\pi/2$, the fibre slices are $S^1_\chi\times S^2$. The circle closes smoothly at $\theta=0$, where $2\sin^2\theta/W\sim\theta^2$; the two-sphere closes at $\theta=\pi/2$, where $\cos^2\theta/W\sim(\pi/2-\theta)^2$. Together these complete the $S^4$ fibre without a boundary. Since $\int_{\Sigma_g}\epsilon_\Sigma=4\pi(g-1)$,
\begin{align}
 \frac1{2\pi}\int_{\Sigma_g}\t dV=2-2g\,,
 \label{eq:CompactF4BundleDegree}
\end{align}
so the connection has integral degree. The warp factor is smooth and nonconstant. The mixed Einstein equation \eqref{eq:iia-Einstein-mixed} holds identically for \eqref{eq:CompactF4Branch}. The flux is
\begin{align}
 \delta_4
 &=2L^3\t d\left[
       \frac{\cos^3\theta}{W}(d\chi+V)\right]
                                      \wedge\epsilon_{S^2}
                                      \nn\\
 &=-2L^3\Bigl[
 \frac{\cos^2\theta\sin\theta(3+\cos^2\theta)}{W^2}
                   d\theta\wedge(d\chi+V)\nn\\
 &\hspace{25mm}
       +\frac{\cos^3\theta}{W}\epsilon_\Sigma
       \Bigr]\wedge\epsilon_{S^2}\,.
 \label{eq:CompactF4Flux}
\end{align}
The three-form potential is local, while $\delta_4$ is globally smooth. Its local expression gives $\t d\delta_4=0$. With orientation $\epsilon_\Sigma\wedge d\theta\wedge(d\chi+V)  \wedge\epsilon_{S^2}$, we have
\begin{align}
 e^{14A/3}\t\star\delta_4
 =-\sqrt2L\Bigl[
 &(3+\cos^2\theta)\epsilon_\Sigma\nn\\
 &+2\sin\theta\cos\theta\,
                 d\theta\wedge(d\chi+V)\Bigr]\,.
 \label{eq:CompactF4Hodge}
\end{align}
Using $\t dV=-\epsilon_\Sigma$, the derivatives of the two terms cancel. Thus $\t d(e^{14A/3}\t\star\delta_4)=0$, as required by \eqref{eq:iia-single-flux-branches}. The $H_3$ equation also holds, since $\delta_4\wedge\delta_4=0$ in six internal dimensions. The internal quantities entering the dilaton equation
\eqref{eq:iia-dilaton-reduced} are
\begin{align}
 e^{8A/3}\delta_4\cdot\delta_4
 &=\frac{2(9+8\cos^2\theta+3\cos^4\theta)}{L^2W^2}\,,
       \nn\\
 \t\nabla^{\t2}e^{4A}
 &=\frac{3e^{2A}}{2L^2}
 \left[\frac{9+8\cos^2\theta+3\cos^4\theta}{W^2}-6\right]\,.
 \label{eq:CompactF4WarpCheck}
\end{align}
Substitution into \eqref{eq:strategy-separated-warp} fixes
\begin{align}
 X_0=\frac1{4L^2}\,.
 \label{eq:CompactF4SeparationConstant}
\end{align}
The internal Einstein equation \eqref{eq:iia-Einstein-internal} becomes
\begin{align}
 &\t R_{mn}-4\t\nabla_m\t\nabla_nA
       -\frac{44}{9}\partial_mA\partial_nA
       -X_0e^{-2A}\t g_{mn}\nn\\
 &\qquad=\frac12e^{2A/3}
 \left[(\delta_4\cdot\delta_4)_{mn}
       -\frac38\t g_{mn}\delta_4\cdot\delta_4\right]\,.
 \label{eq:CompactF4InternalEinstein}
\end{align}
One can check by direct substitution that the internal Einstein equations are satisfied. The warped volume is
\begin{align}
 \mathcal V_A=
 \frac{32\sqrt2\pi^3}{3}(g-1)L^6\,.
 \label{eq:CompactF4WarpedVolume}
\end{align}
With $u=\cos\theta$, the integrated warp equation reduces to
\begin{align}
 \int_0^1du\,
 \frac{u^2(3-4u^2-3u^4)}{(1+u^2)^2}
 =\left[\frac{u^3(1-u^2)}{1+u^2}\right]_0^1=0\,.
 \label{eq:CompactF4Integrability}
\end{align}
Hence \eqref{eq:strategy-integrability} holds without boundary or localised contributions. The nonzero flux periods are
\begin{align}
 \int_{S^4}\delta_4&=-8\pi^2L^3\,,\nn\\
 \int_{\Sigma_g\times S^2|_{\theta=0}}\delta_4
       &=-16\pi^2(g-1)L^3\,.
 \label{eq:CompactF4Periods}
\end{align}
The second period is $2(g-1)$ times the first. We impose flux quantisation using the M-theory uplift \eqref{eq:strategy-F4-uplift}, where the eleven-dimensional four-form field strength is $G_4=\delta_4$. In terms of the eleven-dimensional Planck length $\ell_p$, the quantisation condition gives
\begin{align}
 \frac1{(2\pi\ell_p)^3}\int_{S^4}G_4=-N\,,
 \quad L^3=\pi N\ell_p^3\,,
 \quad N\in\mathbb Z_{>0}\,.
 \label{eq:CompactF4Quantization}
\end{align}
The second period is then automatically quantised. Localised sources are not required in the compact geometry. 

\subsubsection{Isotropic external space solution}

We again use $\alpha_1=\alpha_2=\alpha_3$ and $\sigma=3\alpha_1+12\beta$. The temporal equation in \eqref{eq:strategy-separated-warp} is $-e^{-\sigma+2\beta}\ddot\beta=X_0$. The spatial and $tt$ Einstein equations in \eqref{eq:iia-Einstein-external} give
\begin{align}
 \ddot\beta&=-\frac1{4L^2}e^{3\alpha_1+10\beta}\,,
 \qquad \ddot\alpha_1=18\ddot\beta\,,\nn\\
 0&=\frac32\dot\alpha_1^2
       +18\dot\alpha_1\dot\beta-42\dot\beta^2
       +\frac6{L^2}e^{3\alpha_1+10\beta}\,.
 \label{eq:CompactF4TimeEquations}
\end{align}
A solution of all three equations is
\begin{align}
 \beta&=\frac{\Phi_0}{12}
       -\frac1{32}\ln\cosh(kt)-\frac3{32}kt\,,\nn\\
 \alpha_1&=\alpha_2=\alpha_3
       =18\beta+2kt+\alpha_0\,,\nn\\
 e^{3\alpha_0}
       &=\frac{k^2L^2}{8}e^{-16\Phi_0/3}\,,\quad k>0\,,\quad -\infty<t<\infty\,.
 \label{eq:CompactF4TimeSolution}
\end{align}
The dilaton is
\begin{align}
 \Phi=\Phi_0+\frac14\ln W
       -\frac38\ln\cosh(kt)-\frac98kt\,.
 \label{eq:CompactF4Dilaton}
\end{align}
Here, $\dot\beta=-k[\tanh(kt)+3]/32<0$ throughout.

\subsubsection{Four-dimensional Einstein-frame cosmology}
\begin{figure*}[!t]
  \centering
  \includegraphics[
    width=\textwidth,
    height=0.73\textheight,
    keepaspectratio
  ]{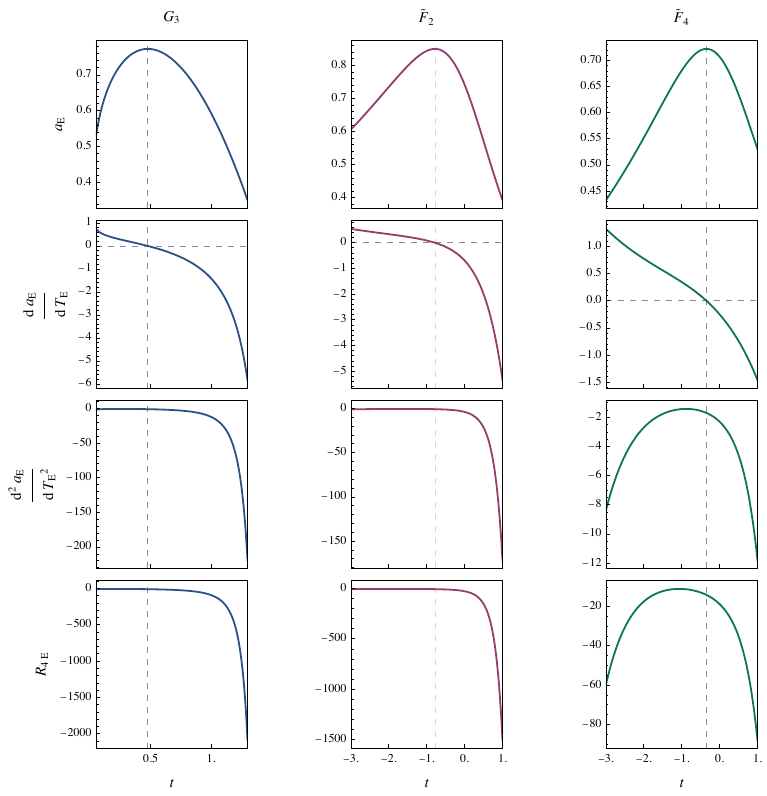}
  \caption{Four-dimensional Einstein-frame cosmology plots for the backgrounds supported by $G_3$ (left), $\tilde F_2$ (centre), and $\tilde F_4$ (right). From top to bottom: scale factor $a_E$, proper-time expansion $da_E/dT_E$, proper-time acceleration $d^2a_E/dT_E^2$, and Ricci scalar $R_{4E}$, each evaluated as a function of coordinate time $t$. We set $L=1$ and $\Phi_0=0$ in all cases, and $k=1$ for the $\tilde F_2$ and $\tilde F_4$ solutions, with $\alpha_0$ fixed by the corresponding background constraint. Vertical dashed lines mark the Einstein-frame turnaround, and the horizontal dashed line in the second row marks when expansion is zero.}
  \label{fig:einstein-frame-cosmology}
\end{figure*}

Here $dT_E=a_E^3dt$, with
\begin{align}
 a_E=e^{\Phi_0+\alpha_0/2}
       [\cosh(kt)]^{-3/8}e^{-kt/8}\,.
 \label{eq:CompactF4EinsteinScale}
\end{align}
The expansion, acceleration, and Ricci scalar are
\begin{align}
 \frac{da_E}{dT_E}
 &=-\frac{k}{8a_E^2}\bigl[3\tanh(kt)+1\bigr]\,,\nn\\
 \frac{d^2a_E}{dT_E^2}
 &=-\frac{k^2}{a_E^5}\left[
       \frac38\operatorname{sech}^2(kt)
       +\frac{[3\tanh(kt)+1]^2}{32}\right]<0\,,\nn\\
 R_{4E}
 &=-\frac{6k^2}{a_E^6}\left[
       \frac38\operatorname{sech}^2(kt)
       +\frac{[3\tanh(kt)+1]^2}{64}\right]<0\,.
 \label{eq:CompactF4Cosmology}
\end{align}
The universe expands for $\tanh(kt)<-1/3$, reaches its maximum size at equality and subsequently contracts. There is no accelerated expansion. The hatted spatial scale increases until $\tanh(kt)=5/9$, after the Einstein-frame turnaround. We show the plots for this in Figure~\ref{fig:einstein-frame-cosmology}. 
\section{Scale separation of the solutions}
\label{sec:scale-separation}

Scale separation requires the characteristic four-dimensional energy scale to be much smaller than the mass of the lightest massive Kaluza--Klein (KK) mode. For a vacuum with characteristic length $L_4$, this means $m_{\mathrm{KK}}L_4\gg1$. For an expanding or contracting cosmology, the corresponding Hubble to KK hierarchy is
\begin{equation}
    \frac{|H_E|}{m_{\mathrm{KK}}}\ll1,
    \qquad
    H_E=\frac{1}{a_E}\frac{da_E}{dT_E}.
    \label{eq:ss-hierarchy}
\end{equation}
Note that both quantities must be evaluated in the same four-dimensional Einstein frame. 

There are arguments which obstruct parametrically scale-separated AdS vacua, provided the KK mass remains tied to the internal curvature scale \cite{GautasonEtAl2016} as discussed in the introduction. These vacuum no-go results rely on assumptions about the matter content, the absence of negative-tension sources and the relation between the internal geometry and the KK mass gap. Andriot, Cribiori and Van Riet \cite{Andriot:2025cyi} demonstrate that rolling solutions can develop scale separation using classical fluxes without orientifolds. In their flux-supported examples, including compactifications on Ricci-flat spaces, the Hubble scale decreases faster than the KK mass, so $H_E/m_{\mathrm{KK}}\to0$ as time increases. Time dependence however does not always guarantee a hierarchy for every rolling solution. For our recollapsing backgrounds, the relevant question is instead whether the ratio becomes small during a finite part of the cosmological evolution. Thus, for the backgrounds constructed here, both the KK masses and the cosmological scales evolve with time. The internal warping enters the eigenvalue problem determining the KK mass gap, while the evolving internal volume determines its time dependence. Moreover, each solution reaches a turnaround at which $H_E=0$, so a small value of $|H_E|/m_{\mathrm{KK}}$ might be possible near the turnaround. In this section, we try to understand this in more detail for each of the three solutions. 

\subsection{The KK mass and the Hubble scale}
\label{subsec:ss-masses}

We first determine the mass generated by internal momentum for a massless, minimally coupled probe scalar $\Psi$. Using the Einstein-frame relation in Section \ref{sec:compact-backgrounds}, the ten-dimensional metric is
\begin{equation}
    ds_{10}^{2}
    =e^{2A-6\beta}ds_{4E}^{2}
     +e^{2\beta}d\tilde{s}_{6}^{2}.
\end{equation}
With the signature conventions of~\ref{app:typeii-conventions}, the scalar action becomes
\begin{align}
 S_{\Psi}
 &=-\frac12\int d^{10}X\,\sqrt{-g_{10}}\,
      g^{MN}\partial_M\Psi\,\partial_N\Psi
      \nn\\
 &=-\frac12\int d^4x\,\sqrt{-g_E}
      \int_{M_6}\t\epsilon\,
      \Bigl[
      e^{2A}g_E^{\mu\nu}\partial_\mu\Psi\,\partial_\nu\Psi
      \nn\\
 &\hspace{35mm}
      +e^{-8\beta}e^{4A}
      \t g^{mn}\partial_m\Psi\,\partial_n\Psi
      \Bigr]\,.
 \label{eq:ss-scalar-action}
\end{align}
Thus the external kinetic term has a time-independent internal weight, while the internal-gradient term carries $e^{-8\beta}$. Expanding $\Psi=\sum_n\psi_n(x)Y_n(y)$ in real eigenfunctions satisfying
\begin{align}
    -\tilde{\nabla}_m
       \left(e^{4A}\tilde{\nabla}^{m}Y_n\right)
       &=\lambda_n e^{2A}Y_n,
       \nonumber\\
    \int_{M_6}e^{2A}Y_nY_r\,\tilde{\epsilon}
       &=\delta_{nr}.
    \label{eq:ss-eigenproblem}
\end{align}
Since the internal manifold is compact and has no boundary, integration by parts gives
\begin{equation}
    \int_{M_6}e^{4A}
       \tilde{\nabla}_mY_n\tilde{\nabla}^{m}Y_r\,
       \tilde{\epsilon}
    =\lambda_n\delta_{nr}.
\end{equation}
Consequently, the canonically normalised four-dimensional modes have action
\begin{equation}
    \begin{split}
    S_{\Psi}
    =-\frac12\sum_n\int d^4x\,\sqrt{-g_E}
      &\left[
        g_E^{\mu\nu}\partial_\mu\psi_n\partial_\nu\psi_n+\lambda_n e^{-8\beta}\psi_n^2
      \right],
    \end{split}
\end{equation}
and KK mass
\begin{equation}
    m_n^2(t)=\lambda_n e^{-8\beta(t)}.
    \label{eq:ss-mode-mass}
\end{equation}
The constant internal eigenfunction has $\lambda_0=0$. For the first nonzero mode,
\begin{equation}
    \lambda_1
    =
    \frac{\displaystyle
      \int_{M_6}e^{4A}|\tilde{\nabla}Y_1|^2\tilde{\epsilon}}
    {\displaystyle
      \int_{M_6}e^{2A}Y_1^2\tilde{\epsilon}}
    =\frac{\mu}{L^2},
    \quad
    m_{\mathrm{KK}}(t)=\frac{\sqrt{\mu}}{L}e^{-4\beta(t)}.
    \label{eq:ss-kk-mass}
\end{equation}
Using Einstein frame relations obtained in Section \ref{sec:compact-backgrounds}, we get
\begin{align}
    dT_E&=e^{\sigma/2+3\beta}dt,
    \nonumber\\
    a_E&=e^{3\beta+\alpha_1/2},
    \nonumber\\
    H_E
    &=e^{-\sigma/2-3\beta}
      \left(3\dot{\beta}+\frac12\dot{\alpha}_1\right),
    \nonumber\\
    \frac{H_E}{m_{\mathrm{KK}}}
    &=\frac{L}{\sqrt{\mu}}e^{\beta-\sigma/2}
      \left(3\dot{\beta}+\frac12\dot{\alpha}_1\right).
    \label{eq:ss-general-hubble}
\end{align}
As in Section \ref{sec:compact-backgrounds}, dots denote derivatives with respect to $t$. The eigenfunctions and KK-mass for all three compact geometries are derived in~\ref{app:kk-spectrum}.

\subsection{The background supported by $H_3$}
\label{subsec:ss-h3}

For the solution of Section \ref{subsec:compact-G3}, $\sigma=2\beta$ and
\begin{equation}
    \begin{split}
    \beta
    &=-\frac{\Phi_0}{4}
     +\frac14\ln\cos(t/L)
     +\frac{1-\sqrt{3}}{8}\ln\tan(t/L),
    \\
    &\quad 0<t<\frac{\pi L}{2}.
    \end{split}
\end{equation}
Substitution into \eqref{eq:ss-kk-mass} and \eqref{eq:ss-general-hubble} gives
\begin{align}
    m_{\mathrm{KK}}
    &=\frac{\sqrt{\mu}\,e^{\Phi_0}}
            {L\cos(t/L)}
      [\tan(t/L)]^{(\sqrt{3}-1)/2},
      \nonumber\\
    H_E
    &=\frac{e^{-4\beta}}{L}
      \frac{\cos(2t/L)-1/\sqrt{3}}{\sin(2t/L)},
      \nonumber\\
    \frac{H_E}{m_{\mathrm{KK}}}
    &=\frac{\cos(2t/L)-1/\sqrt{3}}
            {\sqrt{\mu}\sin(2t/L)}.
    \label{eq:ss-h3-ratio}
\end{align}

From ~\ref{app:kk-h3}, 
\begin{equation}
    \mu
    =\min\left\{
       \underbrace{2+\min(c_1^2,c_1^{-2})}_{\mu_{S^3}},
       \underbrace{\min_a\frac{L^2}{R_a^2}}_{\mu_{T^3}}
      \right\}.
    \label{eq:ss-h3-gap}
\end{equation}
A large torus radius lowers the KK gap but at the same time shrinking the torus raises its contribution only until a sphere mode becomes the lightest mode. Since $2<\mu_{S^3}\leq3$, the full gap cannot be made parametrically large by adjusting $R_a/L$ or $c_1$. At fixed internal shape, changing $L$ or $\Phi_0$ does not change the ratio in \eqref{eq:ss-h3-ratio} at a given $t/L$. Near the big bang, $t/L\to0^+$,
\begin{align}
    m_{\mathrm{KK}}
    &\sim\frac{\sqrt{\mu}\,e^{\Phi_0}}{L}
          (t/L)^{(\sqrt{3}-1)/2}
      \longrightarrow0,
      \nonumber\\
    \frac{H_E}{m_{\mathrm{KK}}}
    &\sim
      \frac{1-1/\sqrt{3}}{2\sqrt{\mu}}\,
      \frac{1}{t/L}
      \longrightarrow+\infty.
    \label{eq:ss-h3-bang}
\end{align}
Thus at early times the solution does not have a hierarchy. The situation changes near the turnaround identified in Section~\ref{subsec:compact-G3},
\begin{equation}
    \frac{t_*}{L}
    =\frac12\arccos\frac{1}{\sqrt{3}}.
\end{equation}
Here $H_E=0$ while $m_{\mathrm{KK}}$ is finite and positive. Expanding \eqref{eq:ss-h3-ratio} around this point gives
\begin{equation}
    \frac{H_E}{m_{\mathrm{KK}}}
    =-\frac{2}{\sqrt{\mu}}\frac{t-t_*}{L}
     +O\left(\frac{(t-t_*)^2}{L^2}\right).
    \label{eq:ss-h3-turnaround}
\end{equation}
For any fixed $0<\epsilon\ll1$, there is therefore a finite interval with $|H_E|/m_{\mathrm{KK}}<\epsilon$. To leading order, its width on either side of the turnaround is
\begin{equation}
    \frac{|t-t_*|}{L}
    <\frac{\epsilon\sqrt{\mu}}{2}.
    \label{eq:ss-h3-window}
\end{equation}
The portion with $t<t_*$ lies within the expanding cosmology. For example, $c_1=2$ and $R_a=L/2$ give $\mu=9/4$.
The exact condition $|H_E|/m_{\mathrm{KK}}<0.1$ then holds
approximately for $0.407<t/L<0.556$, with the turnaround at $t_*/L\simeq0.478$. Thus the small ratio occurs on an interval, not only at the single instant $H_E=0$. We also illustrate this in Figure~\ref{fig:scale-separation-comparison}. 

One can also ask the question of supergravity control in this background. For this $H_3$ solution, 
\begin{equation}
e^{\Phi}=\frac{e^{\Phi_0}}{\sqrt{\cos ^2 \theta+c_1^2 \sin ^2 \theta}} \frac{\tan (t / L)^{(\sqrt{3}-1) / 2}}{\cos (t / L)} .
\end{equation}
For fixed $c_1=O(1)$, a finite regular neighbourhood of turnaround admits
\begin{equation}
e^{\Phi_0} \ll 1, \quad \frac{L^2}{\alpha^{\prime}} \gg 1, \quad R_a \gg \sqrt{\alpha^{\prime}} .
\end{equation}

\subsection{The remaining backgrounds and the duration of the hierarchy}
\label{subsec:ss-general}

The same calculation applies to the $\tilde F_2$ and $\tilde F_4$ backgrounds of Sections~\ref{subsec:compact-F2} and~\ref{subsec:compact-F4}. In both cases $dT_E=a_E^3dt$, and their scale factors give
\begin{align}
    H_E\big|_{\tilde F_2}
    &=-\frac{k}{12a_E^3}
       \left[7\tanh(kt)+\sqrt{21}\right],
       \nonumber\\
    H_E\big|_{\tilde F_4}
    &=-\frac{k}{8a_E^3}
       \left[3\tanh(kt)+1\right].
\end{align}
Using the corresponding $\beta(t)$ and normalisation of $\alpha_0$ from those sections, we obtain
\begin{align}
    \left.\frac{H_E}{m_{\mathrm{KK}}}\right|_{\tilde F_2}
    &=-\frac{7\sinh(kt)+\sqrt{21}\cosh(kt)}
             {2\sqrt{6\mu}},
       \nonumber\\
    \left.\frac{H_E}{m_{\mathrm{KK}}}\right|_{\tilde F_4}
    &=-\frac{3\sinh(kt)+\cosh(kt)}
             {2\sqrt{2\mu}}.
    \label{eq:ss-rr-ratios}
\end{align}
Here $\mu$ is evaluated on the corresponding internal manifold. At the expanding big bang, $t\to-\infty$, both ratios diverge:
\begin{align}
    \left.\frac{H_E}{m_{\mathrm{KK}}}\right|_{\tilde F_2}
    &\sim\frac{7-\sqrt{21}}{4\sqrt{6\mu}}e^{-kt},
       \nonumber\\
    \left.\frac{H_E}{m_{\mathrm{KK}}}\right|_{\tilde F_4}
    &\sim\frac{1}{2\sqrt{2\mu}}e^{-kt}.
\end{align}
Their turnarounds occur at
\begin{equation}
    t_*^{(2)}
    =-\frac1k\operatorname{arctanh}\sqrt{\frac37},
    \qquad
    t_*^{(4)}
    =-\frac1k\operatorname{arctanh}\frac13.
\end{equation}
In terms of these times, \eqref{eq:ss-rr-ratios} can be written exactly as
\begin{align}
    \left.\frac{H_E}{m_{\mathrm{KK}}}\right|_{\tilde F_2}
    &=-\sqrt{\frac{7}{6\mu}}\,
       \sinh\left[k(t-t_*^{(2)})\right],
       \nonumber\\
    \left.\frac{H_E}{m_{\mathrm{KK}}}\right|_{\tilde F_4}
    &=-\frac1{\sqrt{\mu}}\,
       \sinh\left[k(t-t_*^{(4)})\right].
\end{align}
\begin{sloppypar}
Consequently, the exact intervals satisfying $|H_E|/m_{\mathrm{KK}}<\epsilon$ are
\end{sloppypar}
\begin{align}
    k|t-t_*^{(2)}|
    &<\operatorname{arsinh}
       \left(\epsilon\sqrt{\frac{6\mu}{7}}\right),
       \nonumber\\
    k|t-t_*^{(4)}|
    &<\operatorname{arsinh}\left(\epsilon\sqrt{\mu}\right).
    \label{eq:ss-rr-windows}
\end{align}
Each contains a finite expanding portion before the turnaround. The internal gaps are also bounded in these examples. The eigenvalue computations in~\ref{app:kk-f2} and~\ref{app:kk-f4} give
\begin{align}
    \left.\mu\right|_{\tilde F_2}
    &=2,
    \nonumber\\
    \left.\mu\right|_{\tilde F_4}
    &=\lambda_1(\Sigma_g)
      \leq
      \frac{2\left\lfloor(g+3)/2\right\rfloor}{g-1}
      \leq4.
    \label{eq:ss-rr-gap-bounds}
\end{align}
\begin{figure*}[!t]
  \centering
  \includegraphics[width=\textwidth]{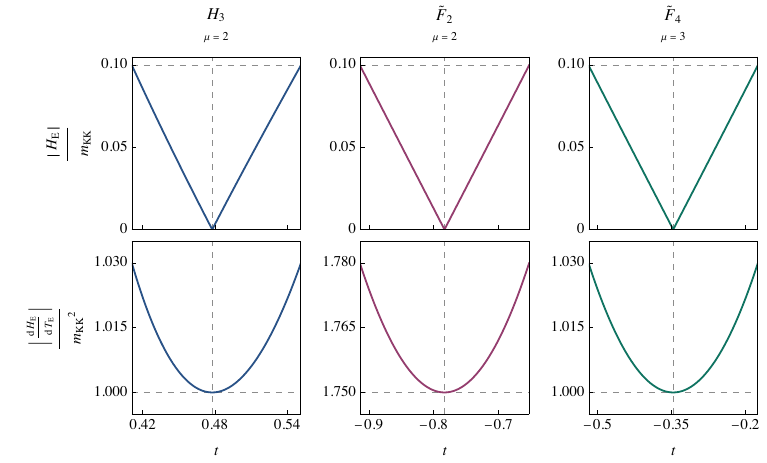}
  \caption{Hubble-to-KK ratio $\mathcal R=|H_E|/m_{\mathrm{KK}}$ (top) and variation ratio $|dH_E/dT_E|/m_{\mathrm{KK}}^2$ (bottom) versus coordinate time $t$ near Einstein-frame turnaround. Columns show $H_3$, $\tilde F_2$ and $\tilde F_4$, with $\mu=2,2,3$, respectively, $L=1$, and $k=1$ in the RR cases. $\lambda_1(\Sigma_g)=3$ was assumed for $\tilde F_4$ within the bound given in \eqref{eq:app-kk-f4-gap}. Vertical dashed lines mark turnarounds, while horizontal lines mark $\mathcal R=0.1$ and the minima $|dH_E/dT_E|/m_{\mathrm{KK}}^2=1,7/4,1$.}
  \label{fig:scale-separation-comparison}
\end{figure*}

Finally, a small hierarchy need not remain small. For the single-flux backgrounds, the Einstein equations in ~\ref{app:detailed-eoms} imply
\begin{equation}
    \frac{dH_E}{dT_E}
    =-3H_E^2
     -\left(3+\frac{\kappa^2}{16}\right)X_0e^{-8\beta}.
    \label{eq:ss-hubble-variation}
\end{equation}
\begin{sloppypar}
Using the values of $\kappa$ and $X_0$ in Sections~\ref{sec:ansatz-eoms} and~\ref{sec:compact-backgrounds} gives
\end{sloppypar}
\begin{equation}
    \frac{|dH_E/dT_E|}{m_{\mathrm{KK}}^2}
    =3\left(\frac{H_E}{m_{\mathrm{KK}}}\right)^2
     +
     \begin{cases}
        2/\mu,   & H_3,\\
        7/(2\mu),& \tilde F_2,\\
        3/\mu,   & \tilde F_4.
     \end{cases}
    \label{eq:ss-derivative-hierarchy}
\end{equation}
At turnaround these quantities are bounded below by $2/3$, $7/4$ and $3/4$, respectively. Thus $|H_E|/m_{\mathrm{KK}}\ll1$ can be arranged near turnaround, but it persists for only a small fraction of time. Figure~\ref{fig:scale-separation-comparison} illustrates the Hubble-to-KK hierarchy near turnaround in the three backgrounds. Although $|H_E|/m_{\mathrm{KK}}$ falls below $0.1$ in the displayed windows, $|dH_E/dT_E|/m_{\mathrm{KK}}^2$ cannot become parametrically small, in agreement with~\eqref{eq:ss-derivative-hierarchy}. A small Hubble-to-KK ratio therefore does not by itself imply a small Hubble-variation ratio.

\section{Supersymmetric or not?}
\label{sec:supersymmetry}

Unbroken supersymmetry therefore requires a nonzero spinor $\epsilon$ for which both the gravitino and dilatino variations vanish. We use the string-frame transformations and write $\epsilon=\epsilon_{\rm s}$. For $m=0$, these are \cite{Becker:2006dvp,Bergshoeff2001}
\begin{align}
 \delta_\epsilon\psi_M
 &=\left(\nabla_M+\frac18H_{MNP}\Gamma^{NP}\Gamma_{11}\right)\epsilon
 \nn\\
 &\quad+\frac{e^\Phi}{8}
 \left(\slashed{\t F}_2\Gamma_M\Gamma_{11}
       +\slashed{\t F}_4\Gamma_M\right)\epsilon=0,
 \label{eq:susy-gravitino}\\
 \delta_\epsilon\lambda
 &=\left(\slashed{\partial}\Phi
       +\frac12\slashed H_3\Gamma_{11}\right)\epsilon
 \nn\\
 &\quad+\frac{e^\Phi}{4}
 \left(3\slashed{\t F}_2\Gamma_{11}
       +\slashed{\t F}_4\right)\epsilon=0.
 \label{eq:susy-dilatino}
\end{align}
Here
\begin{equation}
 \slashed P_p
 =\frac1{p!}P_{M_1\cdots M_p}\Gamma^{M_1\cdots M_p},
 \qquad
 \slashed{\partial}\Phi=\Gamma^M\partial_M\Phi.
\end{equation}
All gamma matrices and covariant derivatives in these variations use the string-frame metric. The metric and supersymmetry parameter are related to their ten-dimensional Einstein-frame counterparts by
\begin{equation}
 g^{\rm s}_{MN}=e^{\Phi/2}g_{MN},
 \qquad
 \epsilon_{\rm s}=e^{\Phi/8}\epsilon_{\rm E}.
 \label{eq:susy-frame-change}
\end{equation}
Together with the corresponding gravitino and dilatino redefinitions, these transformations map the Killing-spinor equations in one frame into those in the other \cite{Hellerman2004}. The redefinitions are invertible wherever the dilaton is finite, so the existence of a nonzero Killing spinor is independent of this choice of frame. We next extract a necessary condition from the dilatino equation. When forming bilinears, the Killing spinor is treated as a commuting Majorana spinor, with
\begin{equation}
 \bar\epsilon=\epsilon^T C,
 \qquad
 C^T=-C,
\end{equation}
where $C$ is the charge-conjugation matrix. For any matrix $M$ that is antisymmetric in its spinor indices,
\begin{equation}
 \epsilon^T M\epsilon
 =(\epsilon^T M\epsilon)^T
 =\epsilon^T M^T\epsilon
 =-\epsilon^T M\epsilon=0.
 \label{eq:susy-antisymmetric-bilinear}
\end{equation}
This implies
\begin{align}
 (C\Gamma^{abc}\Gamma_{11})^T
 &=-C\Gamma^{abc}\Gamma_{11},
 \nn\\
 (C\Gamma^{ab}\Gamma_{11})^T
 &=-C\Gamma^{ab}\Gamma_{11},
 \nn\\
 (C\Gamma^{abcd})^T
 &=-C\Gamma^{abcd},
\end{align}
where $a,b,\ldots$ are orthonormal ten-dimensional indices. Applying \eqref{eq:susy-antisymmetric-bilinear} therefore gives
\begin{align}
 \bar\epsilon\Gamma^{abc}\Gamma_{11}\epsilon&=0,
 \nn\\
 \bar\epsilon\Gamma^{ab}\Gamma_{11}\epsilon&=0,
 \nn\\
 \bar\epsilon\Gamma^{abcd}\epsilon&=0.
 \label{eq:susy-flux-bilinears}
\end{align}
These are precisely the bilinears multiplying $H_3$, $\t F_2$ and $\t F_4$ when \eqref{eq:susy-dilatino} is multiplied by $\bar\epsilon$. Thus every flux term drops out, leaving
\begin{align}
 0=\bar\epsilon\,\delta_\epsilon\lambda
   &=(\bar\epsilon\Gamma^M\epsilon)\partial_M\Phi
    =K^M\partial_M\Phi,
 \nn\\
 K^M&=\bar\epsilon\Gamma^M\epsilon.
 \label{eq:susy-dilaton-current}
\end{align}
This is basically the equation~(4.23) of \cite{Saffin2005}. It states that the dilaton must be invariant along the vector $K$ constructed from the Killing spinor. The time component of this vector provides the obstruction to having supersymmetric backgrounds, as seen below. Choose a real Majorana representation with $C=\Gamma^{\underline0}$, where $\underline0$ denotes the orthonormal time direction. Since $(\Gamma^{\underline0})^2=-1$ in our mostly-plus signature,
\begin{align}
 K^{\underline0}
 &=\bar\epsilon\Gamma^{\underline0}\epsilon
  =\epsilon^T\Gamma^{\underline0}\Gamma^{\underline0}\epsilon
  =-\epsilon^\dagger\epsilon.
 \label{eq:susy-current-norm}
\end{align}
Consequently, $K^{\underline0}$ is nonzero wherever $\epsilon$ is nonzero. To relate this to the coordinate component $K^t$, the string-frame metric gives
\begin{equation}
 g^{\rm s}_{tt}=-e^{\Phi/2+2A+\sigma},
 \qquad
 e_{\rm s}^{\underline0}
 =e^{\Phi/4+A+\sigma/2}dt.
\end{equation}
The orthonormal and coordinate components therefore satisfy
\begin{align}
 K^{\underline0}
 &=e^{\Phi/4+A+\sigma/2}K^t,
 \nn\\
 K^t
 &=e^{-\Phi/4-A-\sigma/2}K^{\underline0}
 \nn\\
 &=-e^{-\Phi/4-A-\sigma/2}\epsilon^\dagger\epsilon.
 \label{eq:susy-current-time}
\end{align}
At a regular point, the exponential factor is finite and strictly positive. Hence $K^t=0$ forces $\epsilon=0$. We will use this together with \eqref{eq:susy-dilaton-current} for each background. We now apply this condition to the three explicit backgrounds. For the $H_3$ background, \eqref{eq:ARKTrial3InternalSolution} gives
\begin{align}
 D&=\cos^2\theta+c_1^2\sin^2\theta,
 \qquad
 A=\frac18\ln D,
 \nn\\
 A'&=\frac{(c_1^2-1)\sin2\theta}{8D}.
 \label{eq:susy-h3-warp}
\end{align}
The differential of $A$ vanishes at $\theta=0$. This is a regular locus where the $\varphi_1$ circle closes. Since $\Phi=-4A-4\beta$, \eqref{eq:ARKTrial3TimeSolution} gives
\begin{equation}
 \left.d\Phi\right|_{\theta=0}
 =-4\dot\beta\,dt
 =\frac{\sqrt3-\cos(2t/L)}{L\sin(2t/L)}\,dt.
 \label{eq:susy-h3-dilaton}
\end{equation}
The coefficient is strictly positive for $0<t<\pi L/2$ and thus non-zero. For the $\t F_2$ background, \eqref{eq:CompactF2Definitions} gives
\begin{align}
 D&=\cos^2\theta+c^2\sin^2\theta,
 \qquad
 A=\frac1{16}\ln D,
 \nn\\
 A'&=\frac{(c^2-1)\sin2\theta}{16D}.
 \label{eq:susy-f2-warp}
\end{align}
Here $dA=0$ at $\theta=\pi/2$, a regular interior value of the polar coordinate. Using $\Phi=4\beta/3+12A$ and \eqref{eq:CompactF2TimeSolution}, we find
\begin{equation}
 \left.d\Phi\right|_{\theta=\pi/2}
 =\frac43\dot\beta\,dt
 =-\frac{k}{12}\bigl[3\tanh(kt)+\sqrt{21}\bigr]dt.
 \label{eq:susy-f2-dilaton}
\end{equation}
This is nonzero for every finite $t$, since $k>0$ and $\sqrt{21}>3$. Finally, for the $\t F_4$ background, \eqref{eq:CompactF4Definitions} gives
\begin{align}
 W&=1+\cos^2\theta,
 \qquad
 A=\frac3{16}\ln W,
 \nn\\
 A'&=-\frac{3\sin2\theta}{16W}.
 \label{eq:susy-f4-warp}
\end{align}
The differential again vanishes at $\theta=0$, where the $\chi$ circle closes smoothly. Using $\Phi=12\beta+4A/3$ and \eqref{eq:CompactF4TimeSolution}, we obtain
\begin{equation}
 \left.d\Phi\right|_{\theta=0}
 =12\dot\beta\,dt
 =-\frac{3k}{8}\bigl[\tanh(kt)+3\bigr]dt.
 \label{eq:susy-f4-dilaton}
\end{equation}
This coefficient is also nonzero for every finite $t$. Therefore, \eqref{eq:susy-dilaton-current} implies $K^t=0$, and \eqref{eq:susy-current-time} forces $\epsilon=0$ at each point. Since vanishing at one regular point implies vanishing throughout, all three solutions break all type IIA supersymmetry.

\section{Discussions}
\label{sec:outlook}
In this article, we have constructed three analytic doubly warped time-dependent backgrounds supported by $H_3$, $\t F_2$ and $\t F_4$, with smooth compact internal manifolds and no localised sources. These examples show that compactness, alongside the absence of localised sources, does not force the internal warp factor to be constant. In particular, the temporal and internal contributions to the dilaton equation can be separated, which paves the way for solving the system. 

All three solutions undergo decelerated expansion followed by contraction in four-dimensional Einstein frame. Although their
internal geometries remain smooth at regular times, this does not remove the cosmological singularities at the temporal endpoints. The KK analysis also illustrates that near turnaround, $|H_E|/m_{\mathrm{KK}}$ becomes arbitrarily small because $H_E$ vanishes at the exact turnaround, while $|dH_E/dT_E|/m_{\mathrm{KK}}^2$ cannot become parametrically small.

These backgrounds are also related to known constructions. For starters, in the string frame, the $H_3$ solution is a combination of a Mueller-type rolling-dilaton cosmology \cite{Mueller:1989in} and marginally deformed $SU(2)$ WZW background \cite{HassanSen1993}. The $\t F_2$ solution arises by circle reduction of an eleven-dimensional vacuum cosmology on $S^3\times S^2\times S^2$, within the framework of time-dependent Einstein-space compactifications \cite{Emparan:2003gg}. Finally, the $\t F_4$ solution retains the internal geometry and flux of the Maldacena--Nu\~nez solution \cite{MaldacenaNunez2001} by replacing its $AdS_5$ factor with a time-dependent Einstein metric of the same Ricci curvature and reducing along a spatial circle, while retaining the internal geometry and four-form flux.

It is natural to ask how the construction changes when the single-flux restriction is relaxed which we plan to investigate in the future. Below we discuss the obstruction for the five-form ansatz and possible extensions towards accelerated expansion.
\subsection{The five-form obstruction}
A time-independent internal five-form, together with its self-dual partner, has no dilaton coupling and gives $E_5=-8$. This restricts the possibility of the integrability condition \eqref{eq:strategy-integrability}. It therefore cannot by itself provide the time-independent source used in the compact construction above. There is also a direct restriction on the $\Omega,\xi$ component of the ansatz in \eqref{eq:CommonAnsatz}. Since $\delta_3$ is real, the purely internal component of $G_3\wedge\bar G_3$ vanishes. For $\Omega\not\equiv0$, \eqref{eq:F5Bianchi06} gives
\begin{equation}
 \t d(e^{-4A}\t\star\t d\xi)=0
 \quad\Longrightarrow\quad
 \int_{M_6}e^{-4A}(\t d\xi\cdot\t d\xi)\t\epsilon=0.
 \label{eq:strategy-F5-obstruction}
\end{equation}
Here $\xi$ is assumed smooth and globally defined. Hence $\t d\xi=0$, and the $[4,1]\oplus[0,5]$ pair vanishes. Thus, in our ansatz, it is not possible to include $\t F_5$ without local sources or without making $\delta_3$ complex. However, we cannot rule out other components of $\t F_5$, like the ones stated in \cite{Kamal:2025qia}, for doubly warped backgrounds. 

\subsection{Accelerating solutions?}
A natural direction is to extend the flux content of our doubly warped backgrounds following \cite{Kamal:2025qia}. A component $[r,s]$ has $r$ external and $s$ internal legs. MN-type components have $r=0$ or $4$, while non-MN-type components have $1\leq r\leq3$. For example, the three-form in \eqref{eq:CommonAnsatz} can be generalised to
\begin{equation}
 G_3=
 \underbrace{\alpha\,\delta_3}_{[0,3]}
 +\underbrace{\zeta_1\wedge\eta_2}_{[1,2]}
 +\underbrace{q\,\h\star\vartheta_1}_{[3,0]}.
 \label{eq:outlook-nonMN-G3}
\end{equation}
The last two terms are non-MN-type because they occupy only part of the four-dimensional spacetime. Their contributions are already retained in the three-form equations of~\ref{app:detailed-eoms} and in the mixed Einstein equation \eqref{eq:iib-Einstein-mixed}. Similarly, the five-form can include the additional self-dual pair
\begin{align}
 \Delta\t F_5
 &=\underbrace{\h\star\gamma_1\wedge\omega_2}_{[3,2]}+
 \underbrace{e^{2\beta-2A}
       \gamma_1\wedge\t\star\omega_2}_{[1,4]},
 \label{eq:outlook-nonMN-F5}
\end{align}
where $\gamma_1$ is an external one-form and $\omega_2$ an internal two-form. Self-duality follows from \eqref{eq:product-hodge}. Together with the $\beta$, non-MN-type fluxes allow contributions of either sign in the traced Einstein equations \cite{Kamal:2025qia}. The Maldacena--Nu\~nez sign argument alone therefore no longer excludes positive four-dimensional Einstein-frame curvature. Finding such backgrounds requires solving the full system alongside integrability for compact manifolds. For asymptotic, spatially flat power-law cosmologies, constraints on stable accelerating scaling solutions at tree level in string theory have also been explored in \cite{VanRiet:2023cca}. However, these arguments do not exclude transient acceleration. We plan to come back to this in future work.

\section*{Acknowledgements}
We deeply thank Ratul Mahanta for detailed discussions on the solutions and for collaboration on earlier stages of this work. The authors were supported by the Research Seed Grant Initiative (RSGI) of BRAC University (BRACURSGI2501). WH acknowledges partial support from the Natural Sciences and Engineering Research Council of Canada (NSERC) via Subatomic Physics Discovery Grant 2026-00042. ARK is funded by the Czech Science Foundation GA\v{C}R grant ``One loop in ten and eleven dimensions" (GA26-22343S). We acknowledge the use of ChatGPT 6 Astra and Claude Opus 5.5 for finding references, cross-checking calculations, and organising the paper. However, the authors take sole responsibility for the material presented.

\appendix
\section{Conventions and Type II equations of motion}
\label{app:typeii-conventions}
\label{app:iib-conventions}

\subsection{Conventions}

We use the Einstein-frame metric and the conventions of \cite{Kamal:2025qia}. The ten-dimensional and external metrics have mostly plus signature. Indices $M,N$, $\mu,\nu$ and $m,n$ refer to the ten-dimensional, four-dimensional and six-dimensional spaces, respectively. Our curvature convention is
\begin{align}
 R^P{}_{QMN}
 &=\partial_M\Gamma^P{}_{NQ}-\partial_N\Gamma^P{}_{MQ}\nn\\
 &\quad+\Gamma^P{}_{MR}\Gamma^R{}_{NQ}
   -\Gamma^P{}_{NR}\Gamma^R{}_{MQ},\nn\\
 R_{QN}&=R^P{}_{QPN},\qquad R_g=g^{MN}R_{MN}.
 \label{eq:curvature-conventions}
\end{align}
Symmetrisation and antisymmetrisation have unit weight. A $p$-form is $P_p=P_{M_1\cdots M_p}dX^{M_1}\wedge\cdots\wedge dX^{M_p}/p!$. For the metric in \eqref{eq:CommonAnsatz}, the orientation is 
\begin{equation}
 \epsilon_{10}=e^{4A+6\beta}\h\epsilon\wedge\t\epsilon,
 \label{eq:volume-conventions}
\end{equation}
where $\h\epsilon$ and $\t\epsilon$ are the positively oriented volume forms of $\h g$ and $\t g$. In $n$ dimensions we define
\begin{equation}
 (\star P_p)_{L_1\cdots L_{n-p}}
 =\frac{\sqrt{|g|}}{p!}
   \epsilon_{L_1\cdots L_{n-p}K_1\cdots K_p}
   P^{K_1\cdots K_p},
 \label{eq:hodge-convention}
\end{equation}
with the output indices first and the positively ordered Levi-Civita symbol equal to $+1$. Thus
\begin{align}
 \star^2P_p&=(-1)^{p(n-p)+s}P_p,\nn\\
 P_p\cdot Q_p&=\frac1{p!}P_{M_1\cdots M_p}Q^{M_1\cdots M_p},\nn\\
 P_p\wedge\star Q_p
 &=(-1)^{p(n-p)}(P_p\cdot Q_p)\epsilon_n,
 \label{eq:star-dot-identities}\\
 d\star df&=-(\nabla^2f)\epsilon_n,\nn\\
 df\wedge\star df&=-(\partial f)^2\epsilon_n
 \qquad(n=4,6,10).
 \label{eq:scalar-hodge-sign}
\end{align}
Here $s$ counts timelike directions. We also use
\begin{equation}
 (P_p\cdot P_p)_{MN}
 =\frac1{(p-1)!}P_{M K_2\cdots K_p}P_N{}^{K_2\cdots K_p}
 \qquad(p\geq1).
 \label{eq:form-tensor-contraction}
\end{equation}
Contractions in this appendix use $g^{MN}$. In the reduced equations, internal dots use $\t g^{mn}$, while external contractions carry hats; in particular $\partial^{\h\mu}=\h g^{\mu\nu}\partial_\nu$ and $\partial^{\t m}=\t g^{mn}\partial_n$. All equations below are bulk equations without localised sources. The string-frame metric is $g^{\rm s}_{MN}=e^{\Phi/2}g_{MN}$.

\subsection{Type IIB}

The field strengths and axiodilaton are
\begin{align}
 H_3&=dB_2,\qquad F_1=dC_0,\nn\\
 F_3&=dC_2,\qquad F_5=dC_4,\nn\\
 \t F_3&=F_3-C_0H_3,\nn\\
 \t F_5&=F_5-\frac12C_2\wedge H_3+\frac12B_2\wedge F_3,\nn\\
 \tau&=C_0+ie^{-\Phi},\nn\\
 G_3&=F_3-\tau H_3=\t F_3-ie^{-\Phi}H_3.
 \label{eq:iib-field-definitions}
\end{align}
In our Hodge convention the Einstein-frame pseudoaction is \cite{Schwarz1983,Kamal:2025qia}
\begin{align}
 S_{\rm IIB}={}&\frac1{2\kappa_{10}^2}\int\Bigl[
 R_g\epsilon_{10}+\frac12d\Phi\wedge\star d\Phi
 +\frac12e^{-\Phi}H_3\wedge\star H_3\nn\\
 &\qquad+\frac12e^{2\Phi}F_1\wedge\star F_1
 +\frac12e^\Phi\t F_3\wedge\star\t F_3\nn\\
 &\qquad+\frac14\t F_5\wedge\star\t F_5\Bigr]\nn\\
 &+\frac1{4\kappa_{10}^2}\int C_4\wedge H_3\wedge F_3.
 \label{eq:iib-pseudoaction}
\end{align}
Five-form self-duality is imposed after variation. The Bianchi identities and real form and scalar equations are
\begin{align}
 dF_1&=dH_3=dF_3=0,\nn\\
 d\t F_3&=H_3\wedge F_1,\qquad
 d\t F_5=H_3\wedge F_3,
 \label{eq:real-bianchi}\\
 0&=d(e^{-\Phi}\star H_3)-F_1\wedge e^\Phi\star\t F_3
       +\t F_5\wedge\t F_3,\nn\\
 0&=d(e^\Phi\star\t F_3)+H_3\wedge\t F_5,\nn\\
 0&=d(e^{2\Phi}\star F_1)+e^\Phi H_3\wedge\star\t F_3,\nn\\
 0&=d\star d\Phi+\frac12e^{-\Phi}H_3\wedge\star H_3
       -e^{2\Phi}F_1\wedge\star F_1\nn\\
 &\quad-\frac12e^\Phi\t F_3\wedge\star\t F_3,\nn\\
 d\star\t F_5&=H_3\wedge F_3,\qquad\star\t F_5=\t F_5.
 \label{eq:real-eoms}
\end{align}
Equivalently, the complex system is
\begin{align}
 d\t F_5&=\frac{i}{2\operatorname{Im}\tau}G_3\wedge\bar G_3,
 \qquad\star\t F_5=\t F_5,\nn\\
 dG_3&=\frac{d\tau\wedge(G_3-\bar G_3)}
                 {2i\operatorname{Im}\tau},\nn\\
 d\star G_3&=-\frac{i}{\operatorname{Im}\tau}
     d\tau\wedge\star\operatorname{Re}G_3+i\t F_5\wedge G_3,\nn\\
 d\star d\tau&=-\frac{i}{\operatorname{Im}\tau}
     d\tau\wedge\star d\tau-\frac{i}{2}G_3\wedge\star G_3.
 \label{eq:iib-complex-system}
\end{align}
The Einstein equation and its trace-reversed form are
\begin{align}
 G_{MN}&=T^1_{MN}+T^3_{MN}+T^5_{MN},\nn\\
 T^1_{MN}&=\frac{\partial_{(M}\tau\partial_{N)}\bar\tau}
 {2(\operatorname{Im}\tau)^2}
 -\frac{g_{MN}\partial_P\tau\partial^P\bar\tau}
 {4(\operatorname{Im}\tau)^2},\nn\\
 T^3_{MN}&=\frac{G_{(M|PQ|}\bar G_{N)}{}^{PQ}}
 {4\operatorname{Im}\tau}
 -\frac{g_{MN}}{4\operatorname{Im}\tau}G_3\cdot\bar G_3,\nn\\
 T^5_{MN}&=\frac1{4\cdot4!}\t F_{MPQRS}\t F_N{}^{PQRS},
 \label{eq:iib-einstein-appendix}\\
 R_{MN}&=\frac{\partial_{(M}\tau\partial_{N)}\bar\tau}
 {2(\operatorname{Im}\tau)^2}
 +\frac{G_{(M|PQ|}\bar G_{N)}{}^{PQ}}
 {4\operatorname{Im}\tau}\nn\\
 &\quad-\frac{g_{MN}}{8\operatorname{Im}\tau}
 G_3\cdot\bar G_3+T^5_{MN}.
 \label{eq:iib-trace-reversed}
\end{align}
Here $G_{MN}=R_{MN}-\frac12g_{MN}R_g$. Self-duality implies $\t F_5\cdot\t F_5=0$, but does not set $T^5_{MN}$ to zero.

\subsection{Type IIA}

We include the Romans parameter $F_0=m$ but the backgrounds considered in the main text have $m=0$. The field strengths are
\begin{align}
 H_3&=dB_2,\qquad F_0=m,\qquad
 \t F_2=dC_1+mB_2,\nn\\
 \t F_4&=dC_3-H_3\wedge C_1+\frac m2 B_2\wedge B_2.
 \label{eq:iia-field-definitions}
\end{align}
The Einstein-frame action is \cite{Polchinski:1998rr, Becker:2006dvp}
\begin{align}
 S_{\rm IIA}={}&\frac1{2\kappa_{10}^2}\int\Bigl[
 R_g\epsilon_{10}+\frac12d\Phi\wedge\star d\Phi
 +\frac12e^{-\Phi}H_3\wedge\star H_3\nn\\
 &\qquad-\frac12e^{5\Phi/2}m^2\epsilon_{10}
 -\frac12e^{3\Phi/2}\t F_2\wedge\star\t F_2\nn\\
 &\qquad-\frac12e^{\Phi/2}\t F_4\wedge\star\t F_4
 \Bigr]+S_{\rm CS},\nn\\
 S_{\rm CS}={}&-\frac1{4\kappa_{10}^2}\int\Bigl[
 B_2\wedge dC_3\wedge dC_3
 +\frac m3 B_2^{\wedge3}\wedge dC_3\nn\\
 &\hspace{36mm}+\frac{m^2}{20}B_2^{\wedge5}\Bigr].
 \label{eq:iia-action}
\end{align}
The relative signs of the odd- and even-degree kinetic terms follow from \eqref{eq:star-dot-identities}. The Bianchi identities are
\begin{align}
 dm&=0,\qquad dH_3=0,\nn\\
 d\t F_2&=mH_3,\qquad d\t F_4=H_3\wedge\t F_2.
 \label{eq:iia-bianchi}
\end{align}
Varying \eqref{eq:iia-action} gives
\begin{align}
 0&=d(e^{3\Phi/2}\star\t F_2)
       +H_3\wedge e^{\Phi/2}\star\t F_4,\nn\\
 0&=d(e^{\Phi/2}\star\t F_4)+H_3\wedge\t F_4,\nn\\
 0&=d(e^{-\Phi}\star H_3)
       +\t F_2\wedge e^{\Phi/2}\star\t F_4\nn\\
 &\quad+m e^{3\Phi/2}\star\t F_2
       +\frac12\t F_4\wedge\t F_4,
 \label{eq:iia-form-eoms}\\
 \nabla^2\Phi&=-\frac12e^{-\Phi}H_3\cdot H_3
       +\frac54m^2e^{5\Phi/2}\nn\\
 &\quad+\frac34e^{3\Phi/2}\t F_2\cdot\t F_2
       +\frac14e^{\Phi/2}\t F_4\cdot\t F_4,
 \label{eq:iia-dilaton}\\
 R_{MN}&=\frac12\partial_M\Phi\partial_N\Phi
       +\frac1{16}m^2e^{5\Phi/2}g_{MN}\nn\\
 &\quad+\frac12e^{-\Phi}\left[
       (H_3\cdot H_3)_{MN}
       -\frac14g_{MN}H_3\cdot H_3\right]\nn\\
 &\quad+\frac12e^{3\Phi/2}\left[
       (\t F_2\cdot\t F_2)_{MN}
       -\frac18g_{MN}\t F_2\cdot\t F_2\right]\nn\\
 &\quad+\frac12e^{\Phi/2}\left[
       (\t F_4\cdot\t F_4)_{MN}
       -\frac38g_{MN}\t F_4\cdot\t F_4\right].
 \label{eq:iia-trace-reversed}
\end{align}

\section{Detailed equations for the doubly warped ansatz}
\label{app:detailed-eoms}

Using~\ref{app:typeii-conventions}, we give the equations following from \eqref{eq:CommonAnsatz}. The geometric formulas hold for $A=A(y)$, $\beta=\beta(x)$ and arbitrary $\h g_{\mu\nu}(x)$. However, the solutions in the main text restrict $\beta$ to be time-dependent only.

\subsection{Common geometric terms}

For an external $p$-form $u_p$ and an internal $q$-form $v_q$,
\begin{align}
 \star(u_p\wedge v_q)
 &=(-1)^{pq}e^{(4-2p)A+(6-2q)\beta}
       \h\star u_p\wedge\t\star v_q,\nn\\
 d(u_p\wedge v_q)
 &=\h d u_p\wedge v_q+(-1)^p u_p\wedge\t d v_q.
 \label{eq:product-hodge}
\end{align}
The scalar Laplacian is
\begin{align}
 \nabla^2 f&=e^{-2A}\left(\h\nabla^{\h2}f
       +6\partial_\mu\beta\partial^{\h\mu}f\right)\nn\\
 &\quad+e^{-2\beta}\left(\t\nabla^{\t2}f
       +4\partial_m A\partial^{\t m}f\right).
 \label{eq:scalar-laplacian-reduced}
\end{align}
The nonzero Ricci blocks and their trace are
\begin{align}
 R_{\mu\nu}
 &=\h R_{\mu\nu}
 -6\left(\h\nabla_\mu\h\nabla_\nu\beta
         +\partial_\mu\beta\partial_\nu\beta\right)\nn\\
 &\quad-e^{2A-2\beta}\h g_{\mu\nu}
   \left(\t\nabla^{\t2}A
         +4\partial_m A\partial^{\t m}A\right),
 \label{eq:Ricci_external}\\
 R_{mn}
 &=\t R_{mn}
 -4\left(\t\nabla_m\t\nabla_n A
         +\partial_m A\partial_n A\right)\nn\\
 &\quad-e^{2\beta-2A}\t g_{mn}
   \left(\h\nabla^{\h2}\beta
         +6\partial_\mu\beta\partial^{\h\mu}\beta\right),
 \label{eq:Ricci_internal}\\
 R_{\mu m}&=8\partial_\mu\beta\partial_m A,
 \label{eq:Ricci_mixed}\\
 R_g&=e^{-2A}\left(\h R-12\h\nabla^{\h2}\beta
                    -42\partial_\mu\beta\partial^{\h\mu}\beta\right)\nn\\
 &\quad+e^{-2\beta}\left(\t R-8\t\nabla^{\t2}A
                    -20\partial_m A\partial^{\t m}A\right).
 \label{eq:Ricci_scalar}
\end{align}
These expressions provide the left-hand sides of the trace-reversed Einstein equations in both theories.

\subsection{Type IIB}
For completeness, we consider the more general IIB three-form ansatz
\[
G_3=\alpha(x)\delta_3(y)
+\zeta_1(x)\wedge\eta_2(y)
+q(y)\hat\star\vartheta_1(x),
\]
together with the metric and five-form ansatz from \eqref{eq:CommonAnsatz}. The three-form ansatz in \eqref{eq:CommonAnsatz}
is recovered by setting $\zeta_1=q=0$. We take $A,\beta,\xi,\Omega$ and $\delta_3,\zeta_1,\vartheta_1$ as real, while $\alpha,\eta_2,q$ may be complex. We retain $\tau(x,y)$ with $\operatorname{Im}\tau>0$. A $[r,s]$-form has $r$ external and $s$ internal legs. 

\subsubsection{Bianchi identities and form equations.} We substitute the ansatzes into \eqref{eq:iib-complex-system}. Since the five-form is self-dual by construction, its equation of motion follows from its Bianchi identity. The independent components of the five-form Bianchi identity are
\begin{align}
 [4,2]:& \quad
 \zeta_1\wedge\h\star\vartheta_1
       \wedge(\bar q\eta_2-q\bar\eta_2)
       = 0,
 \label{eq:F5Bianchi42}\\
 [3,3]:& \quad
 (q\bar\alpha-\bar q\alpha)
       \h\star\vartheta_1\wedge\delta_3
        =0,
 \label{eq:F5Bianchi33}\\
 [1,5]:& \quad
 -e^{-4A}\h d(\Omega e^{4\beta})
       \wedge\t\star\t d\xi
  \nn \\
  & \qquad =\frac{i}{2\operatorname{Im}\tau}
       \zeta_1\wedge
       (\bar\alpha\eta_2-\alpha\bar\eta_2)
       \wedge\delta_3,
 \label{eq:F5Bianchi15}\\
 [0,6]:&\quad
 -\Omega e^{4\beta}
       \t d(e^{-4A}\t\star\t d\xi)=0.
 \label{eq:F5Bianchi06}
\end{align}
The $[2,4]$ component vanishes identically. The purely internal source vanishes because $\delta_3$ is real and $\delta_3\wedge\delta_3=0$. 

The three-form Bianchi identity gives: 
\begin{align}
 [4,0]:\qquad
 &q\,\h d\h\star\vartheta_1
 =\frac{\operatorname{Im}q}{\operatorname{Im}\tau}
 \h d\tau\wedge\h\star\vartheta_1,
 \label{eq:G3Bianchi40}\\
 [3,1]:\qquad
 &\t dq\wedge\h\star\vartheta_1
 =\frac{\operatorname{Im}q}{\operatorname{Im}\tau}
 \t d\tau\wedge\h\star\vartheta_1,
 \label{eq:G3Bianchi31}\\
 [2,2]:\qquad
 &\h d\zeta_1\wedge\eta_2
 =\frac{1}{\operatorname{Im}\tau}
 \h d\tau\wedge\zeta_1\wedge\operatorname{Im}\eta_2,
 \label{eq:G3Bianchi22}\\
 [1,3]:\qquad
 &\begin{aligned}[t]
 \h d\alpha\wedge\delta_3-\zeta_1\wedge\t d\eta_2
 ={}&\frac{\operatorname{Im}\alpha}{\operatorname{Im}\tau}
 \h d\tau\wedge\delta_3\\
 &-\frac{1}{\operatorname{Im}\tau}
 \zeta_1\wedge\t d\tau\wedge\operatorname{Im}\eta_2,
 \end{aligned}
 \label{eq:G3Bianchi13}\\
 [0,4]:\qquad
 &\alpha\,\t d\delta_3
 =\frac{\operatorname{Im}\alpha}{\operatorname{Im}\tau}
 \t d\tau\wedge\delta_3.
 \label{eq:G3Bianchi04}
\end{align}
The three-form equation of motion becomes: 
\begin{align}
 [4,4]:\quad &
 \alpha\h\epsilon\wedge\t d(e^{4A}\t\star\delta_3)
 +e^{2A}\h d(e^{2\beta}\h\star\zeta_1)
       \wedge\t\star\eta_2
 \nn\\
 &=-\frac{i e^{4A}\operatorname{Re}\alpha}
          {\operatorname{Im}\tau}
       \h\epsilon\wedge\t d\tau\wedge\t\star\delta_3\nn\\
 &\quad-\frac{i e^{2A+2\beta}}{\operatorname{Im}\tau}
       \h d\tau\wedge\h\star\zeta_1
       \wedge\t\star\operatorname{Re}\eta_2\nn\\
 &\quad+i\Omega\alpha\h\epsilon\wedge\t d\xi\wedge\delta_3,
 \label{eq:G3EOM44}\\[0.4em]
 [3,5]:\quad &
 -e^{2\beta}\h\star\zeta_1
       \wedge\t d(e^{2A}\t\star\eta_2)\nn\\
 &=\frac{i e^{2A+2\beta}}{\operatorname{Im}\tau}
       \h\star\zeta_1\wedge\t d\tau
       \wedge\t\star\operatorname{Re}\eta_2\nn\\
 &\quad+i\Omega q e^{4\beta-4A}\h\star\vartheta_1
       \wedge\t\star\t d\xi,
 \label{eq:G3EOM35}\\[0.4em]
 [2,6]:\quad &
 q e^{-2A}\h d(e^{6\beta}\vartheta_1)\wedge\t\epsilon\nn\\
 &=-\frac{i\operatorname{Re}q}{\operatorname{Im}\tau}
       e^{-2A+6\beta}\h d\tau\wedge\vartheta_1
       \wedge\t\epsilon.
 \label{eq:G3EOM26}
\end{align}

\subsubsection{Axiodilaton and Einstein equations.} 
Using \eqref{eq:scalar-laplacian-reduced}, the axiodilaton equation is
\begin{align}
 \nabla^2\tau
 &=-\frac{i}{\operatorname{Im}\tau}\left(
 e^{-2A}\partial_\mu\tau\partial^{\h\mu}\tau
 +e^{-2\beta}\partial_m\tau\partial^{\t m}\tau\right)\nn\\
 &\quad-\frac i2\left[
 \alpha^2e^{-6\beta}(\delta_3\cdot\delta_3)\right.\nn\\
 &\hspace{15mm}\left.
 +e^{-2A-4\beta}\zeta_\mu\zeta^{\h\mu}(\eta_2\cdot\eta_2)
 -q^2e^{-6A}\vartheta_\mu\vartheta^{\h\mu}\right].
 \label{eq:tau-components-reduced}
\end{align}
For this ansatz, the mixed five-form contribution $T^5_{\mu m}$ in \eqref{eq:iib-einstein-appendix} vanishes, while the axiodilaton and three-form mixed contributions remain nonzero. We state the components of the trace-reversed Einstein equation \eqref{eq:iib-trace-reversed} below. The external components are
\begin{align}
 R_{\mu\nu}
 &=\frac{\partial_{(\mu}\tau\partial_{\nu)}\bar\tau}
        {2(\operatorname{Im}\tau)^2}
 -\frac{|\alpha|^2e^{2A-6\beta}}
        {8\operatorname{Im}\tau}
       \h g_{\mu\nu}(\delta_3\cdot\delta_3)
 \nn\\
 &\quad+\frac{e^{-4\beta}}{4\operatorname{Im}\tau}
       (\eta_2\cdot\bar\eta_2)
       \left(
       2\zeta_\mu\zeta_\nu
       -\frac12\h g_{\mu\nu}
          \zeta_\rho\zeta^{\h\rho}
       \right)
 \nn\\
 &\quad+\frac{|q|^2e^{-4A}}{4\operatorname{Im}\tau}
       \left(
       2\vartheta_\mu\vartheta_\nu
       -\frac32\h g_{\mu\nu}
          \vartheta_\rho\vartheta^{\h\rho}
       \right)
 \nn\\
 &\quad-\frac{\Omega^2e^{-6A-2\beta}}4
       \h g_{\mu\nu}(\t d\xi\cdot\t d\xi).
 \label{eq:iib-Einstein-external}
\end{align}
The internal components are
\begin{align}
 R_{mn}
 &=\frac{\partial_{(m}\tau\partial_{n)}\bar\tau}
        {2(\operatorname{Im}\tau)^2}
 \nn\\
 &\quad+\frac{|\alpha|^2e^{-4\beta}}
        {4\operatorname{Im}\tau}
       \left[
       \delta_{mpq}\delta_n{}^{\t p\t q}
       -\frac12\t g_{mn}(\delta_3\cdot\delta_3)
       \right]
 \nn\\
 &\quad+\frac{e^{-2A-2\beta}}
        {4\operatorname{Im}\tau}
       \zeta_\rho\zeta^{\h\rho}
       \left[
       2\eta_{(m|p|}\bar\eta_{n)}{}^{\t p}
       -\frac12\t g_{mn}(\eta_2\cdot\bar\eta_2)
       \right]
 \nn\\
 &\quad+\frac{|q|^2e^{2\beta-6A}}
        {8\operatorname{Im}\tau}
       \t g_{mn}\vartheta_\rho\vartheta^{\h\rho}
 \nn\\
 &\quad+\frac{\Omega^2e^{-8A}}4
       \left[
       \t g_{mn}(\t d\xi\cdot\t d\xi)
       -2\partial_m\xi\partial_n\xi
       \right].
 \label{eq:iib-Einstein-internal}
\end{align}
Finally, the mixed components are
\begin{align}
8 \partial_\mu \beta \partial_m A
 &=\frac{
       \partial_\mu\tau\partial_m\bar\tau
       +\partial_m\tau\partial_\mu\bar\tau}
       {4(\operatorname{Im}\tau)^2}
 \nn\\
 &\quad+\frac{e^{-4\beta}}{8\operatorname{Im}\tau}
       \zeta_\mu\delta_m{}^{\t p\t q}
       \left(\bar\alpha\eta_{pq}
             +\alpha\bar\eta_{pq}\right).
 \label{eq:iib-Einstein-mixed}
\end{align}
The geometric expressions for the external and internal Ricci tensors are given in \eqref{eq:Ricci_external}--\eqref{eq:Ricci_internal}.

\subsection{Type IIA}

For the IIA sector we take
\begin{align}
 F_0&=m,\qquad H_3=h_3(y),\qquad\t F_2=\delta_2(y),\nn\\
 \t F_4&=\delta_4(y)+\varpi(x)\h\epsilon,
 \label{eq:iia-flux-ansatz-appendix}
\end{align}
with real forms and arbitrary $\Phi(x,y)$. The two compact R--R backgrounds retain only $\delta_2$ or $\delta_4$, respectively. The spacetime-filling term is included for completeness. 

\subsubsection{Bianchi identities and form equations}
The Bianchi identities reduce to
\begin{equation}
 dm=0,\quad\t d h_3=0,\quad
 \t d\delta_2=mh_3,\quad
 \t d\delta_4=h_3\wedge\delta_2.
 \label{eq:iia-bianchi-reduced}
\end{equation}
The duals entering \eqref{eq:iia-form-eoms} are
\begin{align}
 e^{-\Phi}\star H_3
 &=e^{4A-\Phi}\h\epsilon\wedge\t\star h_3,\nn\\
 e^{3\Phi/2}\star\t F_2
 &=e^{4A+2\beta+3\Phi/2}\h\epsilon\wedge\t\star\delta_2,\nn\\
 e^{\Phi/2}\star\t F_4
 &=e^{4A-2\beta+\Phi/2}\h\epsilon\wedge\t\star\delta_4
   -\varpi e^{-4A+6\beta+\Phi/2}\t\epsilon.
 \label{eq:iia-duals-reduced}
\end{align}
Consequently, the independent form equations are
\begin{align}
 0&=\t d\left(e^{4A+2\beta+3\Phi/2}\t\star\delta_2\right) + e^{4A-2\beta+\Phi/2}h_3\wedge\t\star\delta_4,
 \label{eq:iia-F2-reduced}\\
 0&=\t d\left(e^{4A-2\beta+\Phi/2}\t\star\delta_4\right)
   +\varpi h_3,\nn\\
 0&=\h d\left(\varpi e^{6\beta+\Phi/2}\right),
 \label{eq:iia-F4-reduced}\\
 0&=\t d\left(e^{4A-\Phi}\t\star h_3\right)
   +e^{4A-2\beta+\Phi/2}\delta_2\wedge\t\star\delta_4\nn\\
 &\quad+m e^{4A+2\beta+3\Phi/2}\t\star\delta_2
   +\varpi\delta_4.
 \label{eq:iia-H3-reduced}
\end{align}

\subsubsection{Dilaton and Einstein equations}
Collecting the internal forms and their dilaton couplings as
\begin{align}
 (P_2,P_3,P_4)&=(\delta_2,h_3,\delta_4),\nn\\
 (a_2,a_3,a_4)&=\left(\frac32,-1,\frac12\right).
 \label{eq:iia-internal-flux-notation}
\end{align}
All sums below run over $p=2,3,4$, and contractions of the internal forms $P_p$ use $\t g$. The ten-dimensional four-form norm is $\t F_4\cdot\t F_4=e^{-8\beta}(\delta_4\cdot\delta_4) -e^{-8A}\varpi^2$, with the minus sign coming from the external Lorentzian volume form. The dilaton equation is
\begin{align}
 \nabla^2\Phi
 &=\frac54m^2e^{5\Phi/2}-\frac14\varpi^2 e^{\Phi/2-8A}\nn\\
 &\quad+\frac12\sum_{p=2}^4
 a_p e^{a_p\Phi-2p\beta}(P_p\cdot P_p),
 \label{eq:iia-dilaton-reduced}
\end{align}
where its left-hand side is \eqref{eq:scalar-laplacian-reduced}. The three Einstein blocks are
\begin{align}
 R_{\mu\nu}
 &=\frac12\partial_\mu\Phi\partial_\nu\Phi\nn\\
 &\quad+\frac{\h g_{\mu\nu}}{16}\Bigl[
 m^2 e^{2A+5\Phi/2}-5\varpi^2 e^{\Phi/2-6A}\nn\\
 &\hspace{10mm}
 -\sum_{p=2}^4(p-1)e^{2A+a_p\Phi-2p\beta}(P_p\cdot P_p)
 \Bigr],
 \label{eq:iia-Einstein-external}\\
 R_{mn}
 &=\frac12\partial_m\Phi\partial_n\Phi\nn\\
 &\quad+\frac12\sum_{p=2}^4
 e^{a_p\Phi-2(p-1)\beta}(P_p\cdot P_p)_{mn}\nn\\
 &\quad-\frac{\t g_{mn}}{16}\sum_{p=2}^4
 (p-1)e^{a_p\Phi-2(p-1)\beta}(P_p\cdot P_p)\nn\\
 &\quad+\frac{\t g_{mn}}{16}\left[
 m^2e^{5\Phi/2+2\beta}
 +3\varpi^2e^{\Phi/2+2\beta-8A}\right],
 \label{eq:iia-Einstein-internal}\\
 8\partial_\mu\beta\partial_m A
 &=\frac12\partial_\mu\Phi\partial_m\Phi.
 \label{eq:iia-Einstein-mixed}
\end{align}
The left-hand sides of the first two equations are \eqref{eq:Ricci_external} and \eqref{eq:Ricci_internal}.

For the single-flux cases, set $m=\varpi=h_3=0$ and retain only $\delta_p$, with $p=2$ or $4$. Assuming an additively separated dilaton and nonconstant $A,\beta$, the mixed equation gives $\Phi=\kappa\beta+16A/\kappa$, up to an additive constant. Choosing the cases with time-independent internal equations fixes $a_p\kappa=2(p-1)$. Thus
\begin{align}
 p=2:\quad&\Phi=\frac43\beta+12A,\nn\\
 &\t d\delta_2=0,\qquad
 \t d\left(e^{22A}\t\star\delta_2\right)=0,\nn\\
 p=4:\quad&\Phi=12\beta+\frac43A,\nn\\
 &\t d\delta_4=0,\qquad
 \t d\left(e^{14A/3}\t\star\delta_4\right)=0.
 \label{eq:iia-single-flux-branches}
\end{align}
Here an additive dilaton constant is absorbed into the flux normalisation. These are the two cases used for the compact IIA backgrounds; their remaining equations follow directly from \eqref{eq:iia-dilaton-reduced}--\eqref{eq:iia-Einstein-mixed}.

\section{Internal eigenfunctions and KK mass gaps}
\label{app:kk-spectrum}

In this appendix, we derive the internal eigenfunctions used in Section~\ref{sec:scale-separation}. The relevant operator follows from the minimally coupled scalar action in \eqref{eq:ss-scalar-action}:
\begin{align}
    \mathcal L Y
    &=-e^{-2A}\tilde{\nabla}_m
      \left(e^{4A}\tilde{\nabla}^{m}Y\right)
      \nonumber\\
    &=-\frac{e^{-2A}}{\sqrt{\tilde g}}\,
      \partial_m\left(
      \sqrt{\tilde g}\,e^{4A}\tilde g^{mn}\partial_nY
      \right).
    \label{eq:app-kk-operator}
\end{align}
This is the operator familiar from warped KK reductions \cite{BachasEstes2011}. We use a complex basis for convenience,
\begin{equation}
    \mathcal L Y_I=\lambda_IY_I,
    \qquad
    \int_{M_6}e^{2A}Y_I^*Y_J\,\tilde\epsilon=\delta_{IJ}.
    \label{eq:app-kk-normalization}
\end{equation}
Real and imaginary parts give an equivalent real basis. Since the internal manifolds are smooth, compact and without boundary, integration by parts gives
\begin{equation}
    \int_{M_6}e^{2A}Y_I^*\mathcal L Y_I\,\tilde\epsilon
    =
    \int_{M_6}e^{4A}
      \tilde g^{mn}\partial_mY_I^*\partial_nY_I\,
      \tilde\epsilon
    \geq0.
\end{equation}
The constant mode has $\lambda_0=0$. The first positive eigenvalue determines
\begin{equation}
    \mu=L^2\lambda_1,
    \qquad
    m_{\mathrm{KK}}^2(t)
    =\lambda_1e^{-8\beta(t)}
    =\frac{\mu}{L^2}e^{-8\beta(t)}.
    \label{eq:app-kk-gap-definition}
\end{equation}
Consequently, \eqref{eq:app-kk-gap-definition} gives the mass parameter of the first massive KK mode.
\subsection{The background supported by $H_3$}
\label{app:kk-h3}
For the $H_3$ background, the internal metric is
\begin{equation}
    d\tilde s_6^2
    =e^{2A}\left[
      L^2\left(d\theta^2+f^2d\varphi_1^2
                         +g^2d\varphi_2^2\right)
      +\sum_{a=1}^3R_a^2dz_a^2
      \right],
\end{equation}
where
\begin{align}
    e^{8A}&=\cos^2\theta+c_1^2\sin^2\theta,
    \nonumber\\
    f&=\frac{\sin\theta}
            {\sqrt{\cos^2\theta+c_1^2\sin^2\theta}},
    \quad
    g=\frac{c_1\cos\theta}
            {\sqrt{\cos^2\theta+c_1^2\sin^2\theta}}.
\end{align}
Here $0\leq\theta\leq\pi/2$, while $\varphi_1,\varphi_2$ and $z_a$ have period $2\pi$. The determinant satisfies
\begin{align}
    \sqrt{\tilde g}
    &=e^{6A}L^3R_1R_2R_3fg,
    \nonumber\\
    e^{2A}\sqrt{\tilde g}
    &=c_1L^3R_1R_2R_3\sin\theta\cos\theta,
    \label{eq:app-kk-h3-measure}
\end{align}
where we used $e^{8A}fg=c_1\sin\theta\cos\theta$. Substitution into \eqref{eq:app-kk-operator} gives
\begin{align}
    L^2\mathcal L
    ={}&
    -\frac{1}{\sin\theta\cos\theta}
      \partial_\theta
      \left(\sin\theta\cos\theta\,\partial_\theta\right)
      \nonumber\\
    &-\frac{1}{f^2}\partial_{\varphi_1}^2
     -\frac{1}{g^2}\partial_{\varphi_2}^2
     -\sum_{a=1}^3\frac{L^2}{R_a^2}\partial_{z_a}^2.
\end{align}
Using
\begin{equation}
    \frac{1}{f^2}
    =\frac{1}{\sin^2\theta}+c_1^2-1,
    \qquad
    \frac{1}{g^2}
    =\frac{1}{\cos^2\theta}+c_1^{-2}-1,
\end{equation}
we obtain
\begin{align}
    L^2\mathcal L
    ={}&-\Delta_{S^3}
      -(c_1^2-1)\partial_{\varphi_1}^2
      -(c_1^{-2}-1)\partial_{\varphi_2}^2
      \nonumber\\
    &-\sum_{a=1}^3\frac{L^2}{R_a^2}\partial_{z_a}^2.
    \label{eq:app-kk-h3-operator}
\end{align}
Here $\Delta_{S^3}$ is the scalar Laplacian of the unit round three-sphere in the coordinates
\begin{equation}
    ds_{S^3}^2
    =d\theta^2+\sin^2\theta\,d\varphi_1^2
               +\cos^2\theta\,d\varphi_2^2.
\end{equation}
Periodicity allows the separation
\begin{equation}
    Y=u(\theta)
      e^{in_1\varphi_1+in_2\varphi_2}
      e^{i\sum_aq_az_a},
    \qquad n_1,n_2,q_a\in\mathbb Z.
\end{equation}
For fixed $n_1,n_2$, the round-sphere equation is
\begin{align}
    &-\frac{1}{\sin\theta\cos\theta}
      \frac{d}{d\theta}
      \left(\sin\theta\cos\theta\frac{du}{d\theta}\right)
      \nonumber\\
    &\qquad
      +\left(
        \frac{n_1^2}{\sin^2\theta}
        +\frac{n_2^2}{\cos^2\theta}
       \right)u
      =\ell(\ell+2)u.
    \label{eq:app-kk-h3-radial}
\end{align}
At $\theta=0$, the $\varphi_1$ circle collapses, and smoothness requires $u\sim\theta^{|n_1|}$. At $\theta=\pi/2$, the $\varphi_2$ circle collapses, requiring $u\sim(\pi/2-\theta)^{|n_2|}$. Factoring out these powers reduces \eqref{eq:app-kk-h3-radial} to the Jacobi equation. Regularity at both endpoints gives \cite{LehoucqUzanWeeks2003}
\begin{align}
    u(\theta)
    &=(\sin\theta)^{|n_1|}
      (\cos\theta)^{|n_2|}
      P_r^{(|n_1|,|n_2|)}(\cos2\theta),
      \nonumber\\
    \ell&=|n_1|+|n_2|+2r,
    \qquad r\in\mathbb Z_{\geq0}.
\end{align}
Here and below $P_r^{(a,b)}$ denotes a Jacobi polynomial. The complete eigenfunctions are therefore
\begin{align}
    Y={}&\mathcal N\,
      e^{in_1\varphi_1+in_2\varphi_2+i\sum_aq_az_a}
      (\sin\theta)^{|n_1|}(\cos\theta)^{|n_2|}
      \nonumber\\
    &\times P_r^{(|n_1|,|n_2|)}(\cos2\theta),
    \label{eq:app-kk-h3-modes}
\end{align}
where $\mathcal N$ is fixed by \eqref{eq:app-kk-normalization}. Acting with the remaining terms of \eqref{eq:app-kk-h3-operator} gives
\begin{align}
    L^2\lambda
    ={}&\ell(\ell+2)
      +(c_1^2-1)n_1^2+(c_1^{-2}-1)n_2^2
      \nonumber\\
    &+\sum_{a=1}^3\frac{L^2q_a^2}{R_a^2}.
    \label{eq:app-kk-h3-spectrum}
\end{align}

To identify the lowest sphere excitation, its contribution can be written as
\begin{align}
    L^2\lambda_{S^3}
    ={}&c_1^2n_1^2+c_1^{-2}n_2^2
      +2|n_1n_2|+2(|n_1|+|n_2|)
      \nonumber\\
    &+4r(|n_1|+|n_2|+r+1).
\end{align}
All terms are nonnegative. The lowest nonconstant modes are $\sin\theta\,e^{\pm i\varphi_1}$ and $\cos\theta\,e^{\pm i\varphi_2}$, with eigenvalues $(2+c_1^2)/L^2$ and $(2+c_1^{-2})/L^2$, respectively. The first torus excitations are $e^{\pm iz_a}$, with eigenvalues $1/R_a^2$. Since excitations in the two sectors add, the full gap is
\begin{equation}
    \mu
    =\min\left\{
       2+\min(c_1^2,c_1^{-2}),
       \ \min_a\frac{L^2}{R_a^2}
      \right\}.
    \label{eq:app-kk-h3-gap}
\end{equation}
For finite $c_1>0$, the sphere contribution satisfies $2<\mu_{S^3}\leq3$. Increasing the torus momentum scale therefore raises the full gap only until a sphere mode becomes the lightest excitation.

\subsection{The background supported by $\tilde F_2$}
\label{app:kk-f2}

For the $\tilde F_2$ background, the internal metric is
\begin{align}
    d\tilde s_6^2
    =e^{2A}L^2\bigg[
       \frac12ds_{S^2_1}^2+d\theta^2
       +\frac{\sin^2\theta}{D}(e^\chi)^2
       +ds_{S^2_2}^2
      \bigg],
\end{align}
where
\begin{align}
    e^\chi&=d\chi-\frac p2\cos\theta_1\,d\varphi_1,
    \nonumber\\
    D&=1+\frac{p^2}{2}\sin^2\theta,
    \qquad e^{8A}=\sqrt D.
\end{align}
Both $ds_{S^2_i}^2$ are unit-sphere metrics. The polar angles lie in $[0,\pi]$, all azimuthal angles have period $2\pi$, and $p\in2\mathbb Z\setminus\{0\}$, as specified in the compact solution. We have
\begin{equation}
    e^{2A}\sqrt{\tilde g}
    =\frac{L^6}{2}
      \sin\theta\sin\theta_1\sin\theta_2.
    \label{eq:app-kk-f2-measure}
\end{equation}
The factor $\sqrt D$ from $e^{8A}$ cancels the factor $D^{-1/2}$ from the fibre metric. Define the horizontal sphere operator
\begin{align}
    \Delta_1^{\mathrm H}
    ={}&\frac{1}{\sin\theta_1}
       \partial_{\theta_1}
       \left(\sin\theta_1\,\partial_{\theta_1}\right)
       \nonumber\\
    &+\frac{1}{\sin^2\theta_1}
       \left(
       \partial_{\varphi_1}
       +\frac p2\cos\theta_1\,\partial_\chi
       \right)^2.
\end{align}
Then \eqref{eq:app-kk-operator} becomes
\begin{align}
    L^2\mathcal L
    ={}&-\frac{1}{\sin\theta}
       \partial_\theta\left(\sin\theta\,\partial_\theta\right)
       -\frac{D}{\sin^2\theta}\partial_\chi^2
       \nonumber\\
    &-2\Delta_1^{\mathrm H}-\Delta_{S^2_2}.
    \label{eq:app-kk-f2-operator}
\end{align}

For a Fourier mode $e^{in\chi}$, with $n\in\mathbb Z$, the first sphere carries charge
\begin{equation}
    q_n=\frac{pn}{2}.
\end{equation}
The corresponding monopole harmonics \cite{WuYang1976} satisfy
\begin{align}
    &-\bigg[
       \frac{1}{\sin\theta_1}
       \partial_{\theta_1}
       \left(\sin\theta_1\,\partial_{\theta_1}\right)
       \nonumber\\
    &\hspace{1.2cm}
       +\frac{1}{\sin^2\theta_1}
        \left(\partial_{\varphi_1}
              +iq_n\cos\theta_1\right)^2
      \bigg]\mathcal Y_{jm}^{(q_n)}
      \nonumber\\
    &\hspace{2cm}
      =\left[j(j+1)-q_n^2\right]
        \mathcal Y_{jm}^{(q_n)},
    \label{eq:app-kk-f2-monopole}
\end{align}
where
\begin{equation}
    j=|q_n|,|q_n|+1,\ldots,
    \qquad m=-j,-j+1,\ldots,j.
\end{equation}
These harmonics are sections on the base sphere \cite{WuYang1976}. In the symmetric local coordinates, an explicit expression is
\begin{align}
    \mathcal Y_{jm}^{(q_n)}
    \propto{}&
      e^{im\varphi_1}
      \left(\sin\frac{\theta_1}{2}\right)^{|m+q_n|}
      \left(\cos\frac{\theta_1}{2}\right)^{|m-q_n|}
      \nonumber\\
    &\times
      P_{j-\max(|m|,|q_n|)}^{(|m+q_n|,|m-q_n|)}
      (\cos\theta_1).
    \label{eq:app-kk-f2-monopole-modes}
\end{align}

The fibre-polar equation follows from
\begin{equation}
    \frac{D}{\sin^2\theta}
    =\frac{1}{\sin^2\theta}+\frac{p^2}{2}.
\end{equation}
Its regular eigenfunctions are associated Legendre polynomials,
\begin{equation}
    u(\theta)=P_\ell^{|n|}(\cos\theta),
    \qquad \ell=|n|,|n|+1,\ldots,
\end{equation}
which satisfy
\begin{equation}
    \left[
      -\frac{1}{\sin\theta}\frac{d}{d\theta}
       \left(\sin\theta\frac{d}{d\theta}\right)
      +\frac{n^2}{\sin^2\theta}
    \right]u=\ell(\ell+1)u.
\end{equation}
They vanish as $(\sin\theta)^{|n|}$ at the endpoints where the fibre circle collapses. The complete modes are
\begin{equation}
    Y=\mathcal N\,e^{in\chi}
      P_\ell^{|n|}(\cos\theta)
      \mathcal Y_{jm}^{(q_n)}(\theta_1,\varphi_1)
      Y_{kM}(\theta_2,\varphi_2),
    \label{eq:app-kk-f2-modes}
\end{equation}
where $Y_{kM}$ is an ordinary sphere harmonic, $k\in\mathbb Z_{\geq0}$ and $M=-k,\ldots,k$. All $q_n,j,m$ are integers for the even values of $p$ considered here. The product in \eqref{eq:app-kk-f2-modes} is globally defined on the total space. Indeed, with
\begin{equation}
    \chi_N=\chi-\frac p2\varphi_1,
    \qquad
    \chi_S=\chi+\frac p2\varphi_1,
\end{equation}
its angular phase can be written as
\begin{equation}
    e^{in\chi+im\varphi_1}
    =e^{in\chi_N+i(m+q_n)\varphi_1}
    =e^{in\chi_S+i(m-q_n)\varphi_1}.
\end{equation}
The powers in \eqref{eq:app-kk-f2-monopole-modes} ensure smoothness at the respective base poles. The two expressions agree on their overlap since $\chi_S=\chi_N+p\varphi_1$ and $2q_n=pn$. The fibre terms contribute $\ell(\ell+1)+p^2n^2/2$, the first sphere contributes $2[j(j+1)-q_n^2]$, and the second sphere contributes $k(k+1)$. Since $p^2n^2/2=2q_n^2$, we get after cancelation
\begin{equation}
    L^2\lambda
    =\ell(\ell+1)+2j(j+1)+k(k+1).
    \label{eq:app-kk-f2-spectrum}
\end{equation}
The twisting remains present through $j\geq|pn|/2$. Every term in \eqref{eq:app-kk-f2-spectrum} is nonnegative. The constant mode has $n=\ell=j=k=0$, and the first positive eigenvalue is realised by
\begin{equation}
    Y=\cos\theta
    \qquad\hbox{or}\qquad
    Y=Y_{1M}(\theta_2,\varphi_2).
\end{equation}
Both have $\lambda=2/L^2$. Hence the exact gap is
\begin{equation}
    \mu=2.
    \label{eq:app-kk-f2-gap}
\end{equation}

\subsection{The background supported by $\tilde F_4$}
\label{app:kk-f4}

For the $\tilde F_4$ background, the internal metric is \footnote{The weighted spectral problem on this Maldacena--Nu\~nez geometry has also been studied for spin-two fluctuations in the AdS setting \cite{Chen:2019ydk, DeLuca:2021ojx}.}
\begin{align}
    d\tilde s_6^2
    &=e^{2A}L^2\bigg[
       ds_\Sigma^2+d\theta^2\nn\\
       &\quad
       +\frac{\cos^2\theta}{W}ds_{S^2}^2
       +\frac{2\sin^2\theta}{W}(d\chi+V)^2
      \bigg],
\end{align}
where
\begin{equation}
    W=1+\cos^2\theta,
    \qquad e^{8A}=W^{3/2},
    \qquad dV=-\epsilon_\Sigma.
\end{equation}
The compact base $\Sigma_g$ has Gaussian curvature $-1$, $\chi$ has period $2\pi$, and $0\leq\theta\leq\pi/2$. Writing $ds_{S^2}^2=d\vartheta^2+\sin^2\vartheta\,d\varphi^2$, the measure is
\begin{equation}
    e^{2A}\sqrt{\tilde g}
    =\sqrt{2}L^6\sin\theta\cos^2\theta
      \sqrt{g_\Sigma}\sin\vartheta.
    \label{eq:app-kk-f4-measure}
\end{equation}
The factor $W^{3/2}$ from the warp factor cancels the factor $W^{-3/2}$ from the fibre volume. The Fourier expansion in $\chi$ must respect the bundle over $\Sigma_g$. On an overlap, let
\begin{equation}
    V_\alpha=V_\beta+d\omega_{\alpha\beta},
    \qquad
    \chi_\alpha=\chi_\beta-\omega_{\alpha\beta}.
\end{equation}
A mode $e^{in\chi}\mathcal U_n$ is globally defined when
\begin{equation}
    \mathcal U_{n,\alpha}
    =e^{in\omega_{\alpha\beta}}\mathcal U_{n,\beta}.
\end{equation}
Thus $\mathcal U_n$ is a section of a line bundle over $\Sigma_g$. Introducing its covariant derivative and the corresponding magnetic Laplacian \cite{IengoLi1994,KordyukovTaimanov2022,Charles2024},
\begin{align}
    D_i^{(n)}&=\partial_i-inV_i,
    \nonumber\\
    \mathcal D_n
    &=-\frac{1}{\sqrt{g_\Sigma}}\,
      D_i^{(n)}
      \left(\sqrt{g_\Sigma}g_\Sigma^{ij}D_j^{(n)}\right).
\end{align}
Choose a complete orthonormal basis
\begin{equation}
    \mathcal D_n\mathcal U_{n,a}
    =\Lambda_{n,a}\mathcal U_{n,a}.
    \label{eq:app-kk-f4-base}
\end{equation}
For $n=0$, this is the ordinary scalar Laplacian on $\Sigma_g$, with $\Lambda_{0,0}=0$ and $\Lambda_{0,1}=\lambda_1(\Sigma_g)$. In the local upper-half-plane coordinates used for the background,
\begin{equation}
    ds_\Sigma^2=\frac{du_1^2+du_2^2}{u_2^2},
    \qquad V=-\frac{du_1}{u_2},
\end{equation}
the base equation is explicitly determined by
\begin{equation}
    \mathcal D_n
    =-u_2^2\left(\partial_{u_1}^2+\partial_{u_2}^2\right)
      -2inu_2\partial_{u_1}+n^2.
    \label{eq:app-kk-f4-base-local}
\end{equation}
Its global eigenfunctions are obtained by imposing the identifications of the chosen compact quotient, together with the transition law above. Their detailed form depends on the hyperbolic surface and, for $n\neq0$, the global connection \cite{IengoLi1994}. The local metric alone does not specify these spectral data. We separate the internal modes as
\begin{equation}
    Y=\mathcal N\,e^{in\chi}
      \mathcal U_{n,a}(u_1,u_2)
      Y_{\ell m}(\vartheta,\varphi)\,
      u_{n\ell r}(\theta),
    \label{eq:app-kk-f4-separation}
\end{equation}
where
\begin{equation}
    n\in\mathbb Z,\qquad
    \ell,r\in\mathbb Z_{\geq0},\qquad
    m=-\ell,\ldots,\ell,
\end{equation}
and $-\Delta_{S^2}Y_{\ell m}=\ell(\ell+1)Y_{\ell m}$. The identities
\begin{equation}
    \frac{W}{\cos^2\theta}
    =\frac{1}{\cos^2\theta}+1,
    \qquad
    \frac{W}{2\sin^2\theta}
    =\frac{1}{\sin^2\theta}-\frac12
\end{equation}
reduce the remaining eigenvalue equation to
\begin{align}
    &-\frac{1}{\sin\theta\cos^2\theta}
       \frac{d}{d\theta}
       \left(\sin\theta\cos^2\theta\frac{du}{d\theta}\right)
      +\left[
        \frac{n^2}{\sin^2\theta}
        +\frac{\ell(\ell+1)}{\cos^2\theta}
       \right]u
       \nonumber\\
    &\quad
      =\left[
        L^2\lambda-\Lambda_{n,a}
        -\ell(\ell+1)+\frac{n^2}{2}
       \right]u.
    \label{eq:app-kk-f4-radial}
\end{align}

At $\theta=0$, the $\chi$ circle collapses, so smoothness requires $u\sim(\sin\theta)^{|n|}$. At $\theta=\pi/2$,
the two-sphere collapses, requiring $u\sim(\cos\theta)^\ell$. Factoring out these endpoint powers again gives a Jacobi equation. Its regular solutions are
\begin{equation}
    u_{n\ell r}(\theta)
    =(\sin\theta)^{|n|}
      (\cos\theta)^\ell
      P_r^{(|n|,\ell+1/2)}(\cos2\theta).
    \label{eq:app-kk-f4-radial-modes}
\end{equation}
Substitution into \eqref{eq:app-kk-f4-radial} gives
\begin{align}
    L^2\lambda
    &=\Lambda_{n,a}
      +J(J+3)+\ell(\ell+1)-\frac{n^2}{2},
      \nonumber\\
    J&=|n|+\ell+2r.
    \label{eq:app-kk-f4-spectrum}
\end{align}
Equations \eqref{eq:app-kk-f4-base}, \eqref{eq:app-kk-f4-separation} and \eqref{eq:app-kk-f4-radial-modes} give the complete
separated basis. The fibre dependence is explicit, while the remaining base spectral problem must be solved for the chosen $\Sigma_g$. The first positive eigenvalue can nevertheless be identified without solving the full base spectrum. For $n=0$ and $\ell=r=0$, the modes are pulled back from $\Sigma_g$ and have $L^2\lambda=\Lambda_{0,a}$. The lowest neutral mode with nonconstant fibre dependence has $\ell=1,r=0$ and contributes $6$. The first purely radial excitation has $\ell=0,r=1$ and contributes $10$. For $n\neq0$, the magnetic field has constant magnitude $|n|$. Completing the square gives, for $n>0$,
\begin{align}
    &\int_{\Sigma_g}
      \left|
      \left(D_{\hat 1}^{(n)}-iD_{\hat 2}^{(n)}\right)
      \mathcal U
      \right|^2\epsilon_\Sigma
      \nonumber\\
    &\qquad
      =\int_{\Sigma_g}|D^{(n)}\mathcal U|^2\epsilon_\Sigma
       -n\int_{\Sigma_g}|\mathcal U|^2\epsilon_\Sigma
      \geq0.
\end{align}
Here $\hat 1,\hat 2$ denote an oriented orthonormal frame on $\Sigma_g$. For $n<0$, the opposite sign gives the corresponding identity. Hence
\cite{IengoLi1994}
\begin{equation}
    \Lambda_{n,a}\geq|n|.
\end{equation}
Using $J\geq|n|$, every charged mode obeys
\begin{equation}
    L^2\lambda
    \geq |n|+|n|(|n|+3)-\frac{n^2}{2}
    =\frac{n^2}{2}+4|n|
    \geq\frac92.
    \label{eq:app-kk-f4-charged-bound}
\end{equation}

The Yang--Yau estimate, in its improved genus form, states \cite{YangYau1980,ElSoufiIlias1984}
\begin{equation}
    \lambda_1(\Sigma_g)\operatorname{Area}(\Sigma_g)
    \leq8\pi\left\lfloor\frac{g+3}{2}\right\rfloor.
\end{equation}
Since $\operatorname{Area}(\Sigma_g)=4\pi(g-1)$,
\begin{equation}
    \lambda_1(\Sigma_g)
    \leq
    \frac{2\left\lfloor(g+3)/2\right\rfloor}{g-1}
    \leq4,
    \qquad g\geq2.
\end{equation}
This is below both the charged-sector bound $9/2$ and the first neutral fibre excitation $6$. Therefore the lightest nonconstant mode is pulled back from the base, and the exact gap is
\begin{equation}
    \mu=\lambda_1(\Sigma_g)
    \leq
    \frac{2\left\lfloor(g+3)/2\right\rfloor}{g-1}
    \leq4.
    \label{eq:app-kk-f4-gap}
\end{equation}
Its numerical value ofcourse depends on the chosen hyperbolic metric.

\bibliographystyle{utphys}
\bibliography{biblio}
\end{document}